\documentclass[conference]{IEEEtran}
\IEEEoverridecommandlockouts

\usepackage{cuted} 
\usepackage{newclude}

\makeatletter
\newcommand*{\rom}[1]{\expandafter\@slowromancap\romannumeral #1@}
\makeatother

\usepackage{booktabs}

\usepackage{makecell,multirow}

\usepackage{cite}
\usepackage{amsmath,amssymb,amsfonts}
\usepackage{algorithmic}
\usepackage{graphicx}
\usepackage{caption}
\usepackage{mwe}
\usepackage{textcomp}
\usepackage{xcolor}

\usepackage{upgreek}
\usepackage{color}

\usepackage{float}
\usepackage{multicol}

\usepackage{cleveref}

\usepackage{stmaryrd}
\usepackage[ligature]{semantic}
\mathlig{|->}{\mapsto}
\mathlig{-`}{\rightharpoonup}
\mathlig{~>}{\leadsto}
\mathlig{=>}{\Rightarrow}
\mathlig{|=}{\vDash}
\mathlig{~}{\sim}
\mathlig{>=}{\geq}
\mathlig{<=}{\leq}
\mathlig{<=>}{\Leftrightarrow}
\mathlig{-*}{\sepimp}
\mathlig{->}{\rightarrow}
\mathlig{|[}{\llbracket}
\mathlig{|]}{\rrbracket}
\mathlig{!=}{\neq}

\newtheorem{theorem}{Theorem}[section]
\newtheorem{lemma}{Lemma}[section]
\newtheorem{corollary}{Corollary}[section]
\newtheorem{example}{Example}[section]
\newtheorem{proposition}{Proposition}[section]
\newenvironment{proof}[1][]{
	\textit{Proof #1.}
} {
	\hfill $\square$
}

\newtheorem{fact}{Fact}[section]
\newtheorem{definition}{Definition}[section]

\newcommand{\sepimp}{\mathrel{-\mkern-6mu*}}
\newcommand{\septra}{\mathrel{\sepimp \mkern-15mu^{\lnot}}}
\newcommand{\FV}{\mathrm{FV}}
\newcommand{\dom}{\mathtt{dom}}
\newcommand{\true}{\mathbf{true}}
\newcommand{\false}{\mathbf{false}}
\newcommand{\nil}{\mathbf{nil}}
\newcommand{\emp}{\mathbf{emp}}
\newcommand{\sz}{\mathtt{sz}}
\newcommand{\SeqTerms}{\mathtt{SeqTerms}}

\newcommand{\sx}{s_x}
\newcommand{\sa}{s_{\alpha}}

\newcommand{\FVl}{\FV_{x}}
\newcommand{\SL}{\mathrm{SL}}

\newcommand{\forallx}{\forall_{x}}
\newcommand{\foralla}{\forall_{\alpha}}
\newcommand{\existsx}{\exists_{x}}
\newcommand{\existsa}{\exists_{\alpha}}

\newcommand{\xcf}{x_{c^1}}
\newcommand{\xcs}{x_{c^2}}
\newcommand{\xl}{x_{k}}
\newcommand{\xz}{x_{0}}
\newcommand{\xf}{x_{1}}

\newcommand{\acf}{\alpha_{c^1}}
\newcommand{\acs}{\alpha_{c^2}}
\newcommand{\al}{\alpha_{k}}

\newcommand{\acj}{\alpha_{c^j}}
\newcommand{\acjp}{\alpha_{c^{j'}}}

\def\BibTeX{{\rm B\kern-.05em{\sc i\kern-.025em b}\kern-.08em
    T\kern-.1667em\lower.7ex\hbox{E}\kern-.125emX}}

\allowdisplaybreaks

\begin{document}

\title{A separation logic for sequences \\in pointer programs and its decidability}

\author{
	\IEEEauthorblockN{
		Tianyue Cao\IEEEauthorrefmark{1}, 
		Bowen Zhang\IEEEauthorrefmark{1}, 
		Zhao Jin\IEEEauthorrefmark{3}, 
		Yongzhi Cao\IEEEauthorrefmark{1} and 
		Hanpin Wang\IEEEauthorrefmark{1}\IEEEauthorrefmark{2}
	}
	\IEEEauthorblockA{
		\IEEEauthorrefmark{1}
		Key Laboratory of High Confidence Software Technologies (MOE), \\
		School of Computer Science, Peking University, Beijing, China
	}
	\IEEEauthorblockA{
		\IEEEauthorrefmark{2}
		School of Computer Science and Cyber Engineering, \\
		Guangzhou University, Guangzhou, China
	}
	\IEEEauthorblockA{
		\IEEEauthorrefmark{3}
		School of Computer and Artificial Intelligence, \\
		Zhengzhou University, Zhengzhou, China
	}
	\IEEEauthorblockA{
		tycao@stu.pku.edu.cn, zhangbowen@pku.edu.cn, \\
		jinzhao@zzu.edu.cn, caoyz@pku.edu.cn, whpxhy@pku.edu.cn
	}
}

\maketitle

%
%
%
%

\begin{abstract}
Separation logic and its variants can describe various properties on pointer programs. However, when it comes to properties on sequences, one may find it hard to formalize. To deal with properties on variable-length sequences and multilevel data structures, we propose sequence-heap separation logic which integrates sequences into logical reasoning on heap-manipulated programs. Quantifiers over sequence variables and singleton heap storing sequence (sequence singleton heap) are new members in our logic. Further, we study the satisfiability problem of two fragments. The propositional fragment of sequence-heap separation logic is decidable, and the fragment with 2 alternations on program variables and 1 alternation on sequence variables is undecidable. In addition, we explore boundaries between decidable and undecidable fragments of the logic with prenex normal form. 
\end{abstract}


\section{Introduction}
The classical separation logic proposed by Reynolds\cite{reynolds2002separation} is widely used to describe and verify various properties on heap-manipulated programs. It has evolved in many versions: object-oriented separation logic\cite{parkinson2008separation}, quantitative separation logic \cite{batz2019quantitative}, higher-order separation logic\cite{biering2007bi}, separation logic with inductive predicates\cite{brotherston2014decision}, etc. Its separating conjunction $*$ and separating implication $-*$ provides much succincter expressions on mutable data structures. 

Sequences are also widely used in programs. Many data structures can be abstracted to sequences, including $\mathtt{vector}$ in C++, and $\mathtt{ArrayList}$ in Java. There are many string solvers on sequences, such as CVC4\cite{barrett2011cvc4} and Z3str3\cite{berzish2017z3str3}. Sequence predicate logic\cite{kutsia2004predicate} introduces sequence variables and sequence functions, and allows quantifiers over sequence variables. With sequence variables, sort function can be defined elegantly.

However, to our best knowledge, there is no integral formal systems on logical reasoning of programs with both heap and sequence. The example proposed in \cite{reynolds2002separation} provides a way to formally verify whether a program implements list reversal. The program aims to reverse the sequence $\alpha_0$ in $x$ and put it into $y$. The invariant of this program is written as follows: 
\begin{align}\label{eg:sequence}
	\exists \alpha \exists \beta.\;\bigl((\mathtt{ls}(x, \alpha) * \mathtt{ls}(y, \beta)) \land \alpha_0^{\dag} == \alpha^{\dag} \circ \beta\bigr),
\end{align}
where $\alpha^{\dag}$ denotes reversal of the sequence $\alpha$, and the term $\alpha \circ \beta$ denotes the concatenation of the two sequences $\alpha$ and $\beta$. The predicate $\mathtt{ls}(x,\alpha)$ denotes a singly-linked list starting from $x$, which can be inductively defined as follows:
\begin{equation}\label{eq:def_ls}
	\begin{aligned}
		\mathtt{ls}(x,\varepsilon) &\quad\overset{\triangle}{=}\quad x = \nil, \\
		\mathtt{ls}(x,n \circ \alpha) &\quad\overset{\triangle}{=}\quad \exists y.\; x |-> y \circ n * \mathtt{ls}(y,\alpha).
	\end{aligned}
\end{equation}

The property defined in \Cref{eg:sequence} can be attributed to properties on variable-length sequences in pointer programs. These properties can be found in many commonly-used mutable data structures, such as in stacks, queues, and graphs. Besides, variable-length sequences also appear in multilevel data structures, such as paging on operating systems, and file systems. It is necessary to introduce a logic to describe and verify these kind of properties.

In this paper, we aim to establish a logical foundation for \textbf{variable-length sequences} and \textbf{multilevel data structures} in pointer programs, especially on definition, expressiveness and decidability. We propose sequence-heap separation logic to \textbf{integrate sequences into logical reasoning in pointer programs}. The logic is an extension of both classical separation logic proposed in \cite{reynolds2002separation}, and sequence predicate logic proposed in \cite{kutsia2004predicate}. 

Sequence-heap separation logic can be seen as a fragment of higher-order separation logic proposed in \cite{biering2007bi}. Quantifiers over sequences can be reduced to those over predicates (or over sets). Sequence-heap separation logic is less expressive than higher-order separation logic, but can be used in many scenarios on sequences in pointer programs, and has better deciability results.

In terms of definition for sequence-heap separation logic, we define the heap model as a finite partial mapping from locations to sequences, which is similar to the block model $\mathrm{Heap}_B$ defined in \cite{jin2022reasoning}. The model can be written as $h: \text{Loc} \overset{\mathrm{fin}}{\rightharpoonup} (\text{Loc} \cup \text{Val})^*$, compared to the model $h: \text{Loc} \overset{\mathrm{fin}}{\rightharpoonup} \text{Loc} \cup \text{Val}$ defined in the classical separation logic. 

We also \textbf{define sequence singleton heap} as the term $x |-> \alpha$ to adapt more scenarios on variable-length sequences. The term can be used to denote stack, queue, variadic arguments, trees with variable-length branches, graphs with unbounded out-degree, etc. We take stack as an example. 

In the classical separation logic, one has to describe stack with the inductive predicate $\mathtt{ls}(x,\alpha)$ defined in \Cref{eq:def_ls}, where $x$ and $\alpha$ denote the top pointer and contents of the stack respectively. However, with sequence singleton heap $x |-> \alpha$ defined in sequence-heap separation logic, one can use this formula to denote stack without inductive predicate.

The definition of sequence singleton heap can help us to get a decidable result. We prove that the $\Sigma_1$ fragment with sequence singleton heap is decidable, compared to the undecidable result on $\Sigma_1$ fragment in \cite{demri2018effects} where the list predicate $\mathtt{ls}(x,y)$, separating conjunctions, and separating implications are defined. 

\begin{figure}[!ht]
	\centering
	\includegraphics[width=0.4\textwidth]{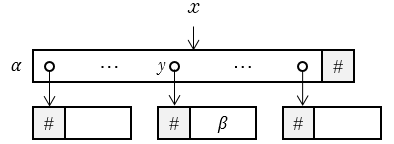}
	\caption{Representation of two-tier data structure}
	\label{fig:two_tier}
\end{figure}

Besides, properly defining the form of $\alpha$ can  help the logic to describe \textbf{multilevel data structures}. For example, if $\alpha$ is of the form $\alpha\#\beta$ (where $\alpha$ denotes the sequence of locations, $\beta$ denotes the sequence of contents, and $\#$ separates these two sequences), one can describe two-tier data structures\cite{jin2022reasoning} in block-based cloud storage systems in the following way.
\begin{align}\label{eq:multi}
	(x |-> \alpha\#\varepsilon) * \forall y\exists \beta.\; \bigl(y \overline{\in} \alpha => (y \hookrightarrow \varepsilon\#\beta)\bigr),
\end{align}
where $x$ denotes the block location, $\alpha$ denotes the sequence of content location, and $\beta$ denotes contents in each location of $\alpha$. The term $y \overline{\in} \alpha$ denotes $y$ appears in $\alpha$, which is defined as follows:
\begin{align*}
	y \overline{\in} \alpha \overset{\triangle}{=} \exists \alpha_1\exists \alpha_2.\; \alpha == \alpha_1 \circ y \circ \alpha_2.
\end{align*}
\Cref{eq:multi} can be illustrated by Fig. (\ref{fig:two_tier}). 



We notice that sequence predicate logic proposed in \cite{kutsia2004predicate} is also capable of describing various properties on variable-length sequences. However, the logic cannot be directly used to describe sequences in pointer programs. sequence-heap separation logic is more than just the combination of sequence predicate logic and separation logic. The logic enriches the composition of the two.

We also notice that the model of sequence-heap separation logic is similar to that of block-based separation logic proposed in \cite{jin2022reasoning}. However, the sequence singleton heap and quantifiers over sequences in sequence-heap separation logic provide more abstraction, and can be used for wider application scenarios, not just for block-based cloud storage systems. 

For example, the combined logic can simplify expressions on describing stack. Besides, sequence predicate logic and separation logic alone cannot describe multilevel data structures shown in \Cref{eq:multi}, but the combined one can do this.

We also study decidability problems of sequence-heap separation logic fragments. This can help us to find algorithms for decidable fragments, as well as to understand expressiveness of the logic intensively. We get the following two main fragments on satisfiability problems. One is a decidable fragment which lays between $\Sigma_1$ fragment of the classical separation logic, and the logic involving list predicate. The other is an undecidable fragment where there are two alternations on quantifiers over program variables, and one alternation on quantifiers over sequence variables.

The paper is organized as follows. In Section \rom{2}, we introduce some basic concepts of separation logic and sequence predicate logic. In Section \rom{3}, we define sequence-heap separation logic and explore its expressiveness. In Section \rom{4}, we prove that the satisfiability problem of $\Sigma_1$ fragment is decidable. In Section \rom{5}, we prove that the satisfiability problem of the fragment with the prenex normal form $\forallx^*\foralla^*\cap \forallx^*\existsx^*\existsa^*$ is undecidable. In addition, we explore boundaries between decidable and undecidable fragments. In Section \rom{6}, we conclude our work, and propose future work on sequence-heap separation logic. 

\textbf{Related work.} Separation logic can express properties on heaps, such as dangling pointers, non-circularity, and memory leakages.\cite{berdine2004decidable} However, the logic itself is proved to be undecidable\cite{brochenin2012almighty}. We investigate the following two categories of fragments: basic fragments without inductive predicates and $\Sigma_1$ fragments with inductive predicates. 

In basic fragments, there are no user-defined predicates or hard-corded predicates. It is known that the validity problem of $\Pi_1$ fragment $\SL(*, -*)$ is decidable\cite{calcagno2001computability}, and that of basic fragment $\SL(\forall, \hookrightarrow)$ with quantifiers and no separation implications is undecidable\cite{calcagno2001computability}. If we restrict the size of the heap satisfying $\varphi_1$ in the formulae $\varphi_1 -* \varphi_2$ to be bounded (namely $n$), then we can get a decidable basic fragment $\SL(*,-*^n,\exists)$\cite{brochenin2012almighty}. To get decidable fragments, one can either restrict quantifier alternations in the formulae of prenex normal form, or number of quantified variables. For quantifier alternations, the fragment $\SL(\exists^*\forall^*)$ with 2 quantifier alternations is decidable\cite{reynolds2017reasoning}, while the fragment $\SL(\exists^*\forall^*\exists^*)$ with 3 is undecidable\cite{reynolds2017reasoning}. For quantified variables, the fragment $1\SL1$ with 1 quantified variable is decidable \cite{demri2017separation}, while the fragment $1\SL2$ with 2 quantified variables is undecidable \cite{demri2015two}. 

%

In $\Sigma_1$ fragments, it contains separation implications (or its variants) and inductive predicates describing data structures. One basic result is that the satisfiability problem of $\Sigma_1$ fragment $\SL(*,-*,\mathtt{ls})$ with lists is undecidable\cite{demri2018effects}. However, if list predicates are restricted to the list with length at most 2, one can get a decidable fragment $\SL(*,-*,\mathtt{reach}_2)$\cite{demri2018effects}. Besides, if we further restrict heap unions in separation model, one can get strong-separation logic\cite{pagel2022strong}. It is proved that the strong-separation fragment $\mathrm{SSL}(*,\septra,\mathtt{ls})$ with list predicates is decidable.

All fragments mentioned above can be found in TABLE \ref{table:decidability}.

\begin{table*}
	\centering
	\normalsize
	\caption{Decidability of separation logic fragments}
	\begin{tabular}{cccc}
		\toprule
		Fragment & Model & \makecell{Key connectives\\and predicates} & Decidability \\
		\midrule
		$\SL(*,-*)$ 
		& \multirow{7}{*}{\makecell{separation\\model}}
		& $\lnot, |->, *,- *$ 
		& decidable
		\\
		$\SL(\forall, \hookrightarrow)$ 
		& 
		& $\lnot, \forall, \hookrightarrow$ 
		& undecidable
		\\
		$\SL(*,-*^n,\exists)$ 
		& 
		& $*\lnot,,-*^n,\exists$ 
		& decidable
		\\
		$\SL(\exists^*\forall^*)$ 
		& 
		& prenex $\exists^*\forall^*$ 
		& decidable
		\\
		$\SL(\exists^*\forall^*\exists^*)$ 
		& 
		& prenex $\exists^*\forall^*\exists^*$ 
		& undecidable
		\\
		$1\SL1$ 
		& 
		& 1 quantified variable 
		& decidable
		\\
		$1\SL2$ 
		& 
		& 2 quantified variables 
		& undecidable
		\\
		\midrule
		$\SL(*,-*,\mathtt{ls})$ 
		& separation model
		& $\lnot, *, -*, \mathtt{ls}(x,y)$ 
		&  undecidable
		\\
		$\SL(*,-*,\mathtt{reach}_2)$ 
		& separation model
		& $\lnot, *, -*, \mathtt{reach}_2(x,y)$ 
		& decidable
		\\
		$\mathrm{SSL}(*,\septra, \mathtt{ls})$ 
		& \makecell{strong-separation\\model} 
		& $\lnot, *, \septra, \mathtt{ls}$
		& decidable
		\\
		\bottomrule
	\end{tabular}
	
	\label{table:decidability}
\end{table*}


Sequence predicate logic is an extension of word equation. Specifically, word equations are $\Sigma_1$ fragment or $\Pi_1$ fragment of sequence predicate logic, where there are only quantifiers over sequence variables. It can express many properties on sequences, such as conjugates, and lexicographic ordering on words\cite{karhumaki2000expressibility}. On the other hand, it cannot express properties such as "the primitiveness" and "the equal length". It is shown that every Boolean combinations of word equations on free semigroup can be reduced to a single word equation\cite{makanin1977problem}.

Different from the classical predicate logic, sequence variables, sequence predicates and sequence functions are defined in sequence predicate logic\cite{kutsia2004predicate}. The problem of whether there is a solution for word equation is decidable\cite{makanin1977problem}. Namely, the truth of $\Sigma_1$ fragment of sequence predicate logic is decidable. Further researches show that the truth of positive formulae (where there is no negations or inequalities) which is of the form  $\forall^*\exists^*$\cite{durnev1995undecidability}, $\exists^*\forall^*$\cite{day2018satisfiability} or $\exists^*\forall^*\exists^*$\cite{vavzenin1983decidability} is shown to be undecidable. To get a decidable fragment, one can restrict the formulae of the form $\exists^*\forall^*$ to the  positive one\cite{vavzenin1983decidability}. 

There are some works which combine separation logic and sequences. When formalizing cloud storage systems, one can construct a modeling language $\textbf{IMDSS}$\cite{jing2017modeling}, and a proof system based on separation logic\cite{jin2022reasoning}. These papers define values of heaps as finite sequences, and describe properties on blocks without using sequence singleton heap. Coq-based proof assistant\cite{zhang2022tool} is implemented based on the logic defined in \cite{jin2022reasoning}.

\section{Preliminaries}
In this section, we introduce some basic concepts on separation logic and sequence predicate logic. 
\subsection{Separation logic}
In this subsection, we mainly focus on definitions and expressiveness of separation logic. 
\begin{definition}[Definition of separation logic]
	The syntax of separation logic\cite{reynolds2000intuitionistic} with 1 records is defined as follows: 
	
	\vspace{-1em}
	\small
	\begin{align*}
		t \quad::=&\quad \nil \mid n \mid x \mid \mathtt{f}^n(t,\dots,t) \\
		p \quad::=&\quad t = t \mid \emp \mid t |-> t \mid \mathtt{P}^n(t,\dots,t) \\
		\varphi \quad::=&\quad p \mid \lnot \varphi \mid \varphi \land \varphi \mid \varphi \lor \varphi \mid \varphi => \varphi \mid \varphi * \varphi \mid \varphi -* \varphi \mid \exists x.\varphi,
	\end{align*}
	\normalsize
	where $n$ denotes constants, $x$ denotes variables, $\mathtt{f}^n$ denotes $n$-ary functions mapping $\mathbb{N}^n$ to $\mathbb{N}$, and $\mathtt{P}^n$ denotes $n$-ary predicates mapping $\mathbb{N}^n$ to $\{0,1\}$. 
	
	The underlying model $\sigma=(s,h)$ of separation logic consists of a stack $s$ and a heap $h$. The set $\mathrm{Loc}$ comprises locations excluding $\nil$. The set $\mathrm{Var}$ comprises symbols of variables, including $x,y,z,\dots$. The set $\mathrm{Val}$ comprises integer values. The model is defined as follows:
	\begin{table}[htbp]
		\centering  
		\label{table1}  
		\tabcolsep=0.5cm
		\renewcommand\arraystretch{1.2}
		\normalsize
		\begin{tabular}{ll}  
			$\text{Loc}, \text{Val} \subseteq \mathbb{N}$ &
			$\nil \in \text{Atoms}$ \\
			$s: \text{Var} -> \text{Loc} \cup \text{Val}$ &
			$h: \text{Loc} \overset{\mathrm{fin}}{\rightharpoonup} (\text{Loc} \cup \text{Val})^*$.
		\end{tabular}
	\end{table}
	\vspace{-0.5em}
	
	The semantics of the term $\mathtt{f}^n(t_1,\dots,t_n)$ can be defined as follows, where $\mathtt{f}_\sigma^n$ is the interpretation of $\mathtt{f}^n$.
	\begin{align*}
		|[\mathtt{f}^n(t_1,\dots,t_n)|]\sigma = \mathtt{f}_\sigma^n(|[t_1|]\sigma,\dots,|[t_n|]\sigma).
	\end{align*}
	
	The semantics of formulae is inductively defined as follows:  
	
	\begin{align*}
		& \sigma |= t_1 = t_2 
		&\text{iff}&\quad 
		|[t_1|]\sigma = |[t_2|]\sigma. \\
		& \sigma |= \emp
		&\text{iff}&\quad
		\mathtt{dom}(h) = \varnothing. \\
		& \sigma |= t_1 |-> t_2
		&\text{iff}&\quad
		|[t_1|]\sigma != \nil \text{ and } \mathtt{dom}(h) = \{|[t_1|]\sigma\} \\
		& & & \quad \text{ and } h(|[t_1|]\sigma) = |[t_2|]\sigma. \\
		& \sigma |= \mathtt{P}^n(t_1,\dots,t_n) 
		&\text{iff}&\quad 
		\mathtt{P}_\sigma^n(|[t_1|]\sigma,\dots,|[t_n|]\sigma) \text{ holds.} \\
		& \sigma |= \lnot \varphi
		&\text{iff}&\quad
		\sigma \nvDash \varphi. \\
		& \sigma |= \varphi_1 \land \varphi_2
		&\text{iff}&\quad
		\sigma |= \varphi_1 \text{ and } \sigma |= \varphi_2. \\
		& \sigma |= \varphi_1 \lor \varphi_2
		&\text{iff}&\quad
		\sigma |= \varphi_1 \text{ or } \sigma |= \varphi_2. \\
		& \sigma |= \varphi_1 => \varphi_2
		&\text{iff}&\quad
		\text{if } \sigma |= \varphi_1 \text{, then } \sigma |= \varphi_2. \\
		& \sigma \;|=\; \varphi_1 * \varphi_2  
		&\text{iff} &\quad 
		\text{there exists heap } h_1 \text{ and } h_2, \\
		& & & \quad \text{such that } \dom(h_1) \cap \dom(h_2) = \varnothing, \\ 
		& & & \quad \text{and } h = h_1 \uplus h_2, \text{ and }  (s,h_1) \;|=\; \varphi_1 , \\
		& & & \quad \text{and } (s,h_2) \;|=\; \varphi_2. \\
		& \sigma \;|=\; \varphi_1 -* \varphi_2 & 
		\text{iff} &
		\quad \text{for all heaps } h_1, \\ 
		& & & \quad \text{if } \dom(h_1) \cap \dom(h) = \varnothing \text{,} \\
		& & & \quad \text{and } (s,h_1) \;|=\; \varphi_1, \\
		& & & \quad \text{then } (s,h_1 \uplus h) \;|=\; \varphi_2. \\
		& \sigma \;|=\; \exists x.\varphi & 
		\text{iff} &
		\quad \text{there exists } x_0 \text{ in Loc} \cup \text{Val,} \\
		& & & \quad \text{such that } (s[x -> x_0], h) \;|=\; \varphi,
	\end{align*}
	where $\mathtt{P}_\sigma^n$ denotes the interpretation of $\mathtt{P}^n$.
\end{definition}

For convenience, we have the following notation\cite{reynolds2000intuitionistic} denoting that $t$ points to $t_1,\dots,t_n$ with some fixed $n$.
\begin{align}\label{eq:singleton}
	t |-> t_1,\dots,t_n \overset{\triangle}{=} (t |-> t_1) * \dots * (t + n - 1 |-> t_n).
\end{align}

In some fragments of separation logic, the heap model is defined as $h: \text{Loc} \overset{\text{fin}}{\rightharpoonup} \text{Val}^n$ with some fixed $n$\cite{reynolds2000intuitionistic}\cite{berdine2005symbolic}. In this case, \Cref{eq:singleton} will no longer hold. The semantics of the notation is defined as follows: 
\begin{align*}
	& \sigma |= t |-> t_1,\dots,t_n &\text{iff}\quad& |[t|]\sigma != \nil \text{ and } \mathtt{dom}(h) = \{|[t|]\sigma\}, \\
	& & & \text{ and } h(|[t|]\sigma) = (|[t_1|]\sigma,\dots,|[t_n|]\sigma).
\end{align*}

We notice that in either separation logic or one of its fragments (symbolic heap\cite{berdine2004decidable}\cite{berdine2005symbolic}, strong-separation logic\cite{pagel2022strong}, etc.), the singleton heap is defined with some fixed number $n$. It may not be suitable for describing properties on blocks, or graphs where each node is of unbounded out-degrees. 

The paper \cite{jin2022reasoning} introduces sequence into separation logic to describe properties on block-based cloud storage systems. The model of the logic is defined as a quintuple of the form $(s_F, s_B, s_V, h_B, h_V)$, where $s_F, s_B, s_V$ are assignments of file variables, block variables and location variables respectively, $h_B, h_V$ are heaps of blocks and heaps of values respectively. The file stack $s_F$ is a mapping from file variables to a sequence of block addresses, which is denoted as $s_F \overset{\triangle}{=} \text{FVar} -> \text{BLoc}^*$, and $h_B$ is a finite partial mapping from blocks to sequences of locations, which is denoted as $h_B \overset{\triangle}{=} \text{BLoc} \overset{\text{fin}}{\rightharpoonup} \text{Loc}^*$. Although $s_F$ and $h_B$ both range over sequences, the logic does not introduce sequence variables, sequence functions, or sequence singleton heap. The logic visits these sequences by existential quantifiers on block variables, and by equality over values of block variables.

\subsection{Word equations and sequence predicate logic}
Word equation can be seen as a special case of sequence predicate logic. It can be defined as follows: 
\begin{definition}[Word equation]
	Suppose $X^*$ is a free semigroup satisfying $X = \{n_1,\dots,n_k\}$ with unknowns $\alpha_1,\dots,\alpha_m \in X^*$, where $n_1,\dots,n_k$ are generators. Word equation is an equality of the form:
	\begin{align*}
		U(n_1,\dots,n_k,\alpha_1,\dots,\alpha_m) == V(n_1,\dots,n_k,\alpha_1,\dots,\alpha_m),
	\end{align*}
	where $==$ is the equality between two concatenations of sequences. If there is a solution for $U == V$, then the word equation is satisfied.
\end{definition}

The definition of word equation can be enhanced with Boolean combination, as is shown in \Cref{def:Boolcomb}. 

\begin{definition}[Boolean combination of word equations]\label{def:Boolcomb}
	Boolean combination of word equations is defined as follows: 
	\begin{align*}
		t &\quad::=\quad n \mid \alpha \mid t \circ t \\
		\varphi &\quad::=\quad t == t \mid \lnot \varphi \mid \varphi \land \varphi \mid \varphi \lor \varphi,
	\end{align*}
	where $n$ denotes constants, $\alpha$ denotes sequence variables, and $t_1 \circ t_2$ denotes the concatentation of two sequences $t_1$ and $t_2$.
\end{definition}

Word equation is capable of describing many properties on sequences. Every Boolean combinations of word equations can be reduced to a single word equation followed by \Cref{thm:Boolean comb}\cite{karhumaki2000expressibility}.

\begin{theorem}\label{thm:Boolean comb}
	For every Boolean combinations $\varphi$ defined above, there exists a single word equation $T(\varphi) \overset{\triangle}{=} U == V$, which is equivalent to $\varphi$. The proof sketch of the theorem can be found in Appendix. 
\end{theorem}

If word equation is enhanced with sequence function, and quantifiers over variables and sequences, then one can get the following definition of sequence predicate logic.
\begin{definition}[Sequence predicate logic]\label{def:seqPredLogic}
	The syntax of sequence predicate logic\cite{kutsia2004predicate} can be defined as follows: 
	\begin{align*}
		t &\quad::=&& n \mid x \mid \alpha \mid t \circ t \mid \mathtt{f}(t,\dots,t) \mid \overline{\mathtt{f}}(t,\dots,t) \\
		\varphi &\quad::=&& t == t \mid \mathtt{P}(t,\dots,t) \mid \lnot \varphi \mid \varphi \land \varphi \mid \varphi \lor \varphi \\
		&&& \mid \varphi => \varphi \mid \exists x.\varphi \mid \exists \alpha.\varphi.
	\end{align*}
	The variable $x$ denotes individual variables, and $\alpha$ denotes sequence variables. The arity of $\mathtt{f}$, $\overline{\mathtt{f}}$, and $\mathtt{P}$ can be either fixed or flexible. When it is fixed, there is no sequence variables on parameters. When it is flexible, it may contain sequence variables. 
	
	The structure of the logic is defined as $(\mathrm{Val},I)$, where $\mathrm{Val}$ is the domain, and $I$ is the interpretation on individual and sequence constants, functions and predicates. The assignment of the logic can be defined as $(\sx,\sa)$, where $\sx: \mathrm{IVar} -> \mathrm{Val}$, $\sa: \mathrm{SVar} -> \mathrm{Val}^*$, and $\mathrm{IVar} = \{x,y,z,\dots\}$, $\mathrm{SVar} = \{\alpha,\beta,\gamma,\dots\}$. 
	
	In sequence predicate logic, $t_1 \circ t_2$ denotes the concatenation between two sequences $t_1$ and $t_2$, $t_1 == t_2$ denotes that two equations $t_1$ and $t_2$ are equal. Due to space limitations, we do not list the semantics of the logic. The details can be found in \cite{kutsia2004predicate}\cite{kutsia2007solving}.
\end{definition}

\subsection{Expressiveness of sequence predicate logic}
We recall properties which can be expressed by sequence predicate logic. They are useful for describing properties with sequence-heap separation logic.

\noindent\textbf{Length}. Sequence predicate logic can be used to get the length of a sequence. The function $\mathtt{length}(\alpha)$ denotes the length of the sequence $\alpha$, which can be inductively defined as follows:
\begin{align*}
	\mathtt{length}&(\varepsilon) \overset{\triangle}{=} 0 \\
	\mathtt{length}&(n \circ \alpha) \overset{\triangle}{=} 1 + \mathtt{length}(\alpha).
\end{align*}

For convenience, we define $|\alpha|$ to denote $\mathtt{length}(\alpha)$.

\noindent\textbf{Statistics}. Sequence predicate logic can be used to get occurrences of specific terms. The function $\mathtt{find}(\alpha, x)$ denotes the occurrences of $x$ in sequence $\alpha$, which can be inductively defined as follows:
\begin{align*}
	\mathtt{find}&(\varepsilon, x) \overset{\triangle}{=} 0 & \\
	\mathtt{find}&(n \circ \alpha, x) \overset{\triangle}{=} 1 + \mathtt{find}(\alpha, x), & n = x \\
	\mathtt{find}&(n \circ \alpha, x) \overset{\triangle}{=} \mathtt{find}(\alpha, x), & n \neq x
\end{align*}

For convenience, we define $|\alpha|_{x}$ to denote $\mathtt{find}(\alpha, x)$. 

\noindent\textbf{Lookups}. The predicate $\mathtt{eq}(x_1, \alpha, x_2)$ denotes the $x_2$-th item of sequence $\alpha$ is $x_1$, which can be expressed as follows: 
\begin{align*}
	\mathtt{eq}(x_1, \alpha, x_2) \overset{\triangle}{=} \exists \alpha_1 \exists \alpha_2.\; \alpha == \alpha_1 \circ x_1 \circ \alpha_2 \land |\alpha_1 \circ x_1| = x_2.
\end{align*}

Note that $\alpha(x)$ does not appear as a term in \Cref{def:seqPredLogic}. If so, consider the case where $x$ exceeds the length of sequence $\alpha$. A new value $\perp$ should be introduced to denote illegal term, which may complicate the logic. We take $x_1 = \alpha(x_2)$ to denote $\mathtt{eq}(x_1,\alpha,x_2)$ for convenience:
\begin{align*}
	x_1 = \alpha(x_2) \overset{\triangle}{=} \mathtt{eq}(x_1,\alpha,x_2).
\end{align*}


Other properties can be expressed with lookups. We list some of them below. 

The relationships between items and sequences can be defined as $x \;\bar{\in}\; \alpha$, which means $x$ can be found in the sequence $\alpha$: 
\begin{align*}
	x \;\bar{\in}\; \alpha \overset{\triangle}{=} \exists x_1.\; x = \alpha(x_1).
\end{align*}

The $x_1$-th item of the sequence $\alpha_1$ is equal to the $x_2$-th item of the sequence $\alpha_2$: 
\begin{equation*}
	\mathtt{eq}(\alpha_1, x_1, \alpha_2, x_2) \overset{\triangle}{=} \exists x_3.\; x_3 = \alpha_1(x_1) \land x_3 = \alpha_2(x_2).
\end{equation*}

\noindent\textbf{Unary property}. All items in the sequence $\alpha$ satisfies the unary property $\mathtt{P}^1(x)$:
\begin{align*}
	\forall x \forall \alpha_1 \forall \alpha_2.\; \alpha == \alpha_1 \circ x \circ \alpha_2 => \mathtt{P}^1(x).
\end{align*}

\noindent\textbf{Binary property}. Each two items with different indices in the sequence $\alpha$ satisfies the binary property $\mathtt{P}^2(x_1,x_2)$ :
\begin{align*}
	\forall x_1 \forall x_2 \forall \alpha_1 \forall \alpha_2 \forall \alpha_3.\;& \alpha == \alpha_1 \circ x_1 \circ \alpha_1 \circ x_2 \circ \alpha_3 \\
	& => \mathtt{P}^2(x_1,x_2).
\end{align*}

Increment and set-like defined bellow are binary properties.

\noindent\textbf{Increment}. The sequence $\alpha$ is strictly increasing: 
\begin{align*}
	\mathtt{Inc}(\alpha) \overset{\triangle}{=} \;& \forall x_1 \forall x_2 \forall \alpha_1 \forall \alpha_2 \forall \alpha_3.\\  
	& \alpha == \alpha_1 \circ x_1 \circ \alpha_2 \circ x_2 \circ \alpha_3 => x_1 < x_2.
\end{align*}

\noindent\textbf{Set-like}. Each two items in sequence $\alpha$ are distinct:
\begin{align*}
	\mathtt{Diff}(\alpha) \overset{\triangle}{=} \;& \forall x_1 \forall x_2 \forall \alpha_1 \forall \alpha_2 \forall \alpha_3. \\
	& \alpha == \alpha_1 \circ x_1 \circ \alpha_1 \circ x_2 \circ \alpha_3 => x_1 != x_2.
\end{align*}



\noindent\textbf{Segment}. The sequence $\alpha_2$ is the consecutive sub-sequence of $\alpha_1$: 
\begin{equation*}
	\alpha_2 \;\in\; \mathtt{Seg}(\alpha_1)  \overset{\triangle}{=} \exists \alpha_3 \exists \alpha_4.\\ \alpha_1 == \alpha_3 \circ \alpha_2 \circ \alpha_4.
\end{equation*}


\noindent\textbf{Truncation}. The sequence $\alpha_2$ is the consecutive sub-sequence of $\alpha_1$ with indices ranging over $[x_1, x_2)$ :
\begin{align*}
	&\alpha_2 ==\; 
	\mathtt{Trunc}(\alpha_1,x_1,x_2) \\
	\overset{\triangle}{=} \quad& \exists \alpha_3 \exists \alpha_4.\; \alpha_1 == \alpha_3 \circ \alpha_2 \circ \alpha_4 \\
	& \hspace{4em} \land |\alpha_3| = x_1 - 1 \land |\alpha_3 \circ \alpha_2| = x_2 - 1.
\end{align*}

\section{sequence-heap separation logic}
sequence-heap separation logic is an extension of both classical separation logic and sequence predicate logic. It can describe properties on variable-length sequences, and multilevel data structures. Some properties cannot be expressed easily by the  classical separation logic or sequence predicate logic alone. 

In this section, we present the definition and expressiveness of sequence-heap separation logic (SeqSL). 

\subsection{Definition of sequence-heap separation logic}
\begin{definition}[Syntax of sequence-heap separation logic]\label{def:syntaxSeqSL}
The syntax of the logic can be defined as follows: 
\begin{align*}
	t_x &\quad::= \quad \mathbf{nil} \mid \# \mid n \mid x \\
	t_{\alpha} &\quad::= \quad \varepsilon \mid t_x \mid \alpha \mid t_{\alpha} \circ t_{\alpha} \\
	\varphi &\quad::= \quad t_x = t_x \mid t_{\alpha} == t_{\alpha} \mid \lnot \varphi \mid \varphi \land \varphi \mid \varphi \lor \varphi \mid \varphi => \varphi \\
	& \hspace{3.5em} \mid \emp \mid t_x |-> t_\alpha \mid \varphi * \varphi \mid \varphi -* \varphi \mid \existsx x.\varphi \mid \existsa \alpha.\varphi.
\end{align*}
\end{definition}
Note that other functions and predicates can be defined in SeqSL following the definition in \cite{kutsia2004predicate}, such as comparison between sequence, and picking up an item from a sequence. Most commonly-used predicates and functions can be expressed by the logic defined above. 

\begin{definition}[Model of sequence-heap separation logic]\label{def:modelSeqSL}
There are three types of variables in the signature of SeqSL: program variables $\text{PVar} = \{x, y, z, \dots\}$, sequence variables $\text{SVar} = \{\alpha, \beta, \gamma, \dots\}$. These sets are all countable. The model $\sigma = (\sx,\sa,h)$ can be defined as follows: 

\begin{table}[H]
	\centering  
	\label{table1}  
	\tabcolsep=0.5cm
	\renewcommand\arraystretch{1.2}
	\normalsize
	\begin{tabular}{ll}  
		$\text{Loc}, \text{Val} \subseteq \mathbb{N}$ &
		$\nil, \# \in \text{Atoms}$ \\
		$\text{Loc} \cap \text{Atoms} = \varnothing$ &
		$\sx: \text{PVar} -> \text{Loc} \cup \text{Val}$ \\
		$\sa: \text{SVar} -> (\text{Loc} \cup \text{Val})^*$ &
		$h: \text{Loc} \overset{\text{fin}}{\rightharpoonup} (\text{Loc} \cup \text{Val})^*$,
	\end{tabular}
\end{table}
\vspace{-0.5em}
where Loc is the set of locations, Val is the set of values, Atoms is the set of reserved words, $\sx$ denotes the stack, $\sa$ denotes the assignment on sequence variables, and $h$ denotes the heap which stores sequence of locations, values, or atoms.  
\end{definition}


\begin{definition}[Semantics of terms]\label{def:smtsTerms}
The symbol $t_x$ denotes \textbf{individual term}, whose value ranges over Val. The denotational semantics of the term can be inductively defined as follows: 
\begin{align*}
	|[&\nil|]\sigma = \nil &
	|[&\#|]\sigma = \# \\
	|[&n|]\sigma = n &
	|[&x|]\sigma = \sx(x),
\end{align*}
where $\nil, \#$ in Atoms cannot be used as a location, and $\#$ is used to describe multilevel data structures. 

The symbol $t_{\alpha}$ denotes \textbf{sequence term}, whose value ranges over $(\text{Loc} \cup \text{Val})^*$ . The denotational semantics of the term can be inductively defined as follows:
\begin{align*}
	|[&\varepsilon|]\sigma = \varepsilon &
	|[&t_x|]\sigma \text{ is defined above} \\
	|[&\alpha|]\sigma = \sa(\alpha) &
	|[&t_{\alpha_1} \circ t_{\alpha_2}|]\sigma = |[t_{\alpha_1}|]\sigma \circ |[t_{\alpha_2}|]\sigma, 
\end{align*}
where $|[t_{\alpha_1}|]\sigma \circ |[t_{\alpha_2}|]\sigma$ denotes the concatenation of two sequences $|[t_{\alpha_1}|]\sigma$ and $|[t_{\alpha_2}|]\sigma$. 
\end{definition}

\begin{definition}[Semantics of sequence-heap separation logic]\label{def:semanticsSeqSL}
In SeqSL, the indices of sequence items start from 1. We set $\text{Values} = \text{Val} \; \cup\;  \text{Loc}$ for convenience. The semantics of SeqSL can be defined as follows: 
\begin{align*}
	& \sigma \;|=\; t_{x_1} = t_{x_2} &
	\text{iff} &
	\quad |[t_{x_1}|]\sigma = |[t_{x_2}|]\sigma. \\
	& \sigma \;|=\; t_{\alpha_1} == t_{\alpha_2} &
	\text{iff} &
	\quad |[t_{\alpha_1}|]\sigma == |[t_{\alpha_2}|]\sigma. \\
	& \sigma \;|=\; \lnot \varphi &
	\text{iff} &
	\quad \sigma \;\nvDash\; \varphi. \\
	& \sigma \;|=\; \varphi_1 \land \varphi_2 & 
	\text{iff} &
	\quad \sigma \;|=\; \varphi_1 \text {, and } \sigma \;|=\; \varphi_2. \\
	& \sigma \;|=\; \varphi_1 \lor \varphi_2 & 
	\text{iff} &
	\quad \sigma \;|=\; \varphi_1 \text {, or } \sigma \;|=\; \varphi_2. \\
	& \sigma \;|=\; \varphi_1 => \varphi_2 & 
	\text{iff} &
	\quad \text{if } \sigma \;|=\; \varphi_1 \text {, then } \sigma \;|=\; \varphi_2. \\
	& \sigma \;|=\; \textbf{emp} & 
	\text{iff} &
	\quad \dom(h) = \varnothing. \\
	& \sigma \;|=\; t_x |-> t_\alpha & 
	\text{iff} &
	\quad |[t_x|]\sigma \notin \mathrm{Atom}, \; \dom(h) =  \{|[t_x|]\sigma\},\\
	& & & \quad \text{ and } h(|[t_x|]\sigma) = |[t_{\alpha}|]\sigma \\
	& \sigma \;|=\; \varphi_1 * \varphi_2 & 
	\text{iff} &
	\quad \text{there exist heap } h_1 \text{ and } h_2, \\
	& & & \quad \text{such that } \dom(h_1) \cap \dom(h_2) = \varnothing \;, \\ 
	& & & \quad \text{and } h = h_1 \uplus h_2,\; (\sx,\sa,h_1) \;|=\; \varphi_1, \\
	& & & \quad \text{and } (\sx,\sa,h_2) \;|=\; \varphi_2. \\
	& \sigma \;|=\; \varphi_1 -* \varphi_2 & 
	\text{iff} &
	\quad \text{for all heaps } h_1, \\
	& & & \quad \text{if } \dom(h_1) \cap \dom(h) = \varnothing \text{ ,} \\
	& & & \quad \text{and } (\sx,\sa,h_1) \;|=\; \varphi_1, \\
	& & & \quad \text{then } (\sx,\sa,h_1 \uplus h) \;|=\; \varphi_2. \\
	& \sigma \;|=\; \existsx x.\varphi & 
	\text{iff} &
	\quad \text{there exists } x_0 \text{ in Values,} \\
	& & & \quad \text{such that } (\sx[x -> x_0], \sa, h) \;|=\; \varphi. \\
	& \sigma \;|=\; \existsa \alpha.\varphi &
	\text{iff} &
	\quad \text{there exists } \alpha_0 \text{ in Values}^*, \\
	& & & \quad \text{such that } (\sx, \sa[\alpha -> \alpha_0], h) \;|=\; \varphi,
\end{align*}
where $\sx[x -> x_0]$ and $\sa[\alpha -> \alpha_0]$ denote substitution of program variable and sequence variable respectively. They are defined as follows. 
\begin{align*}
	\sx[x -> x_0](x') = \begin{cases}
		\sx(x_0), & x' = x \\
		\sx(x),   & \text{otherwise}
	\end{cases}
\end{align*}
\begin{align*}
	\sa[\alpha -> \alpha_0](\alpha') = \begin{cases}
		\sa(\alpha_0), & \alpha' == \alpha \\
		\sa(\alpha),   & \text{otherwise}
	\end{cases}
\end{align*}	
\end{definition}



For convenience, we write $\exists x.\varphi$ to denote $\existsx x.\varphi$, and $\exists \alpha.\varphi$ to denote $\existsa \alpha.\varphi$.

SeqSL can express properties which is widely used in logic reasoning on pointer programs. We list some of them. These notations will be used in this paper. 
\begin{align*}
	& \text{septraction} &&\quad \varphi_1 \septra \varphi_2 \overset{\triangle}{=} \lnot (\varphi_1 -* \lnot \varphi_2) \\
	& \text{universal quantifier} &&\quad \forall \alpha.\varphi \overset{\triangle}{=} \lnot \exists \alpha.\; \lnot \varphi \\
	& \alpha \text{ is stored in } x &&\quad x \hookrightarrow \alpha \overset{\triangle}{=} x |-> \alpha * \true \\
	& x \text{ has been allocated } &&\quad \mathtt{alloc}(x) \overset{\triangle}{=} \exists \alpha.\; x \hookrightarrow \alpha \\
	& \alpha_2 \text{ is stored in } \alpha_1(x_1) && \quad \alpha(x_1) \hookrightarrow \alpha_2 \\
	&&& \hspace{-0.5em}\overset{\triangle}{=}\; \exists x.\; x = \alpha_1(x_1) \land x \hookrightarrow \alpha_2.
\end{align*}

\subsection{Expressiveness of sequence-heap separation logic}
In this subsection, we list few examples to show expressiveness of SeqSL. We divide examples into 2 categories: properties on variable-length sequence in programs, and properties on multilevel data structures.

Similar to the classical separation logic, sequence-heap separation logic does not make differences between locations and values in formulae. To adjust more scenarios in practice, we use the atom $\#$ to manually separate locations and values. The sequence singleton heap can thus be written as $x |-> \alpha_l \circ \# \circ \alpha_x$, where $\alpha_l$ denotes sequence of locations, and $\alpha_x$ denotes sequence of values. 

\subsubsection{Properties on variable-length sequences in programs}

we list some properties on graphs to show the expressiveness of SeqSL on describing variable-length sequences in programs. These properties include: out-degree and reachability. 



\noindent\textbf{Out-degree}. The out-degree of node $x$ can be denoted as $\mathtt{Outdeg}(x)$ . The formula $\mathtt{Outdeg}(x) = n$ means the out-degree of node $x$ is $n_x$. It is defined as follows:
\begin{equation}\label{eq:out_degree}
\begin{aligned}
	&\mathtt{Outdeg}(x) = n_x \\
	\overset{\triangle}{=} \quad& \exists \alpha_l \exists \alpha_x.\;x \hookrightarrow \alpha_l \circ \# \circ \alpha_x \land |\alpha_l| = n_x.
\end{aligned}
\end{equation}


Note that the reserved word $\textbf{nil}$ might be in $\alpha_l$. It does not affect the truth of the definition, because $\textbf{nil}$ can be viewed as location 0. Its out-degree is 0, and in-degree is not necessarily be 0. 

%

\noindent\textbf{Reachability}. First, we define one-step reachability $x_1 \leadsto x_2$ , which means there exists an edge from $x_1$ to $x_2$. 
\begin{equation*}
	x_1 \leadsto x_2 \overset{\triangle}{=} \exists \alpha_{l_1} \exists \alpha_{l_2} \exists \alpha_x.\; l_1 \hookrightarrow \alpha_{l_1} \circ l_2 \circ \alpha_{l_2} \circ \# \circ \alpha_x.
\end{equation*}

Then we define $\mathtt{reach}^n(x_1,x_2)$, which means there exists a path from $x_1$ to $x_2$. The length of the path is $n$ ($n >= 0$). 
\begin{align*}
	\mathtt{reach}^0(x_1,x_2) &\quad\overset{\triangle}{=}\quad  x_1 = x_2  \\
	\mathtt{reach}^{n+1}(x_1,x_2) &\quad\overset{\triangle}{=}\quad \exists x_3.\; x_1 \leadsto x_3 * \mathtt{reach}^{n}(x_3,x_2)
\end{align*}

Finally we define $\mathtt{reach}(x_1,x_2)$, which means there exists a path from $x_1$ to $x_2$, whose length is larger or equal than 0. 
\begin{align*}
	\mathtt{reach}(x_1,x_2) \overset{\triangle}{=} \exists n.\; n >= 0 \land \mathtt{reach}^{n}(x_1,x_2).
\end{align*}

Note that $\mathtt{reach}(x_1,x_2)$ defined in SeqSL is different from $\mathtt{ls}(x_1,x_2)$ defined in the separation logic fragments on shape analysis. The former can describe reachability on graphs without restrictions on out-degree of each nodes, while the latter can only describe reachability on linked lists.

\subsubsection{Properties on multilevel data structures}

we list some properties on two kinds of storage systems to show expressiveness of SeqSL on describing multilevel data structures.

Different from SeqSL, the model $\sigma = (s_x,s_\alpha, h)$ of SeqSL describing storage systems should be interpreted in disk, not in memory.

\noindent\textbf{Properties on Windows storage systems}. The structure of Windows storage systems based on trees can be expressed by SeqSL. The indices $\alpha_l$ of files are stored in the location $l$. The file $\alpha_x$ is stored in the location $l_1$. 
\begin{equation}\label{eq:multi_level}
\begin{aligned}
	& l \hookrightarrow \alpha_l \circ \# \circ \varepsilon \;* \\
	& \forall l_1 \exists \alpha_x.\; (l_1 \overline{\in} \alpha_l => l_1 \hookrightarrow \varepsilon \circ \# \circ \alpha_{x}).
\end{aligned}
\end{equation}
Note that the sub-formula $\forall l_1 \exists \alpha_x.\; (l_1 \overline{\in} \alpha_l => l_1 \hookrightarrow \varepsilon \circ \# \circ \alpha_x)$ can not be replaced by $\forall x.\; \alpha_l(x) \hookrightarrow \varepsilon \circ \# \circ \alpha_x$. The reason is that, $\alpha_l$ is a \textit{finite} sequence. When $x$ exceeds the length of $\alpha_l$, it falsifies $\forall x.\; \alpha_l(x) \hookrightarrow \varepsilon \circ \# \circ \alpha_x$. So we have to restrict $l_1$ with $l_1 \overline{\in} \alpha_l$. 

We apply $\#$ in \Cref{eq:multi_level} to describe two-tier data structures. The first tier merely stores locations $\alpha_l$, while the second tier merely stores data $\alpha_x$. These two different forms of contents prevent the location $l_1$ in the second tier from pointing to the location $l$ in the first tier.

With \Cref{eq:multi_level} in hand, we can describe properties of files. We list two of them. 

A file can be found by its index $x_1$: 
\begin{align*}
	\exists \alpha_x.\; \alpha_l(x_1) \hookrightarrow \varepsilon \circ \# \circ \alpha_x.
\end{align*}

The content $x_1$ can be found in the file $\alpha_x$: 
\begin{align*}
	\exists x_2.\; x_1 = \alpha_x(x_2).
\end{align*}

In addition, we can apply $x |-> \alpha_l \circ \#^n \circ \alpha_x$ to describe n-th tier of multilevel data structure. This is useful for describing multilevel pointers in C++, multilevel paging, multilevel addressing, etc. 

\noindent\textbf{Properties on block-based cloud storage systems}. SeqSL can be used to describe blocks in cloud storage systems. Generally speaking, the sequence $\alpha$ in the formula $x \hookrightarrow \alpha$ can be viewed as a block. With a relative address $l$, the content $\alpha(l)$ can be found. 

For convenience, we define the predicate $\mathtt{IncIndex}(\alpha_0, n)$, which means $\alpha_0$ with  length $n+1$ is strictly increasing. The first item of $\alpha$ is 1, and the last is $n + 1$. 

\vspace{-1em}
\begin{align*}
	\mathtt{IncIndex}(\alpha_0, n) \;\overset{\triangle}{=} \; & \mathtt{Inc}(\alpha_0) \land |\alpha_0| = n + 1 \\
	& \land 1 = \alpha_0(1) \land n + 1 = \alpha_0(n+1).
\end{align*}

A big file $\alpha$ stored in location $l$ is divided into $n$ smaller blocks and is stored in disk. The indices $\alpha_l$ of the file are stored in location $l$. The function $\mathtt{Trunc}(\alpha,x_1,x_2)$ can be used to locate these blocks. The sequence variable $\alpha_0$ in the following formula represents the sequence of division points, which should make the predicate  $\mathtt{IncIndex}(\alpha_l,n)$ to be true. The length of each block $\alpha_x$ can be additionally fixed to 64 by tahe formula $|\alpha_x| = 64$. 
\begin{align*}
	& \exists \alpha_0.\; \Bigl( \mathtt{IncIndex}(\alpha_l,n) \land l \hookrightarrow \alpha_l \circ \# \circ \varepsilon \\
	& * \bigl(\forall l_1 \exists x.\; (l_1 = \alpha_l(x) \land \alpha_x == \mathtt{Trunc}(\alpha, \alpha_0(x), \alpha_0(x+1))) \\
	& => l_1 \hookrightarrow \varepsilon \circ \# \circ \alpha_x \Bigr).
\end{align*}

\begin{figure}[h]
	\centering
	\includegraphics[width=0.5\textwidth]{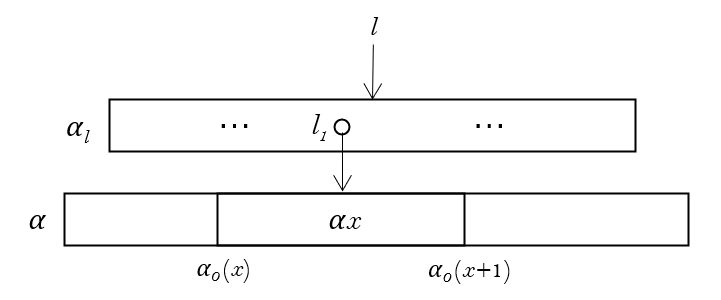}
	\caption{Structure of block-based cloud storage system}
	\label{fig:bcss}
\end{figure}

The formula defined above can be illustrated by Fig. (\ref{fig:bcss}).

\section{Decidable fragments}

In this section, we consider a decidable $\Sigma_1$ fragment of SeqSL. It contains sequence singleton heap, separating conjunction, separating implication, equality and concatenation on sequences. We present the following main result of this section.

\begin{theorem}[Decidable fragment of SeqSL\label{decidable_fragment_of_SeqSL}]\label{theorem:decidable_fragment}
	The satisfiability problem of the language is decidable when it is restricted as follows: 
	\begin{align*}
		t_x \quad::=& \quad \nil \mid \# \mid x \\
		t_{\alpha} \quad::=& \quad \varepsilon \mid t_x \mid \alpha \mid t_{\alpha} \circ t_{\alpha}  \\
		\varphi \quad::=& \quad t_x = t_x \mid t_\alpha == t_\alpha  \mid \false \mid \varphi => \varphi \mid \mathbf{emp} \\
		& \quad \mid x |-> t_{\alpha} \mid \varphi * \varphi \mid \varphi -* \varphi.
	\end{align*}
	
	The model of the fragment is the same as that of SeqSL, which is defined in \Cref{def:modelSeqSL}.
\end{theorem}

We call the above fragment the $\Sigma_1$ fragment of separation logic \;(PSeqSL, where the alphabet 'P' means propositional). PSeqSL is a non-trivial fragment for the following reasons: 
\begin{itemize}
	\item The expressiveness of PSeqSL is stronger than that of the $\Sigma_1$ fragment of the classical separation logic proposed in \cite{reynolds2002separation}, because the former can express properties on variable-length sequences and on multilevel data structures.
	\item The heap stores variable-length sequences in PSeqSL. The models of a formula in PSeqSL may have more possibilities than that of a formula in the classical separation logic. 
	\item Sequences and heap operations in the fragment are not independent. Their satisfiability may affect each other. For instance, to decide whether the formula $x_1 |-> \alpha_1 \land x_1 |-> \alpha_2$ is true, we need to consider both conditions on heap, and the implicit condition $\alpha_1 == \alpha_2$ on sequences. To decide whether the formula $(x_1 |-> \alpha_1 * x_2 |-> \alpha_2) \land (x_1 |-> \alpha_3 * x_2 |-> \alpha_4) \land x_1 != x_2$ is true, we need to consider the implicit condition $\alpha_1 == \alpha_3 \land \alpha_2 == \alpha_4$. 
	\item PSeqSL can describe some of the formulae consisting of existential quantifiers in SeqSL. The property $\mathtt{alloc}(x) \overset{\triangle}{=} \exists \alpha.\; x \hookrightarrow \alpha$ in SeqSL can be expressed in PSeqSL as: 
	\begin{align*}
		\mathtt{alloc}(x) \overset{\triangle}{=} (x |-> \nil) -* \false.
	\end{align*}
\end{itemize}

We prove \Cref{decidable_fragment_of_SeqSL} following these three steps:  
\begin{enumerate}
	\item prove the following problem is decidable: given the formula $\varphi$ in PSeqSL, and a part of the model $\sx, h$, decide whether there exists $\sa$, such that $(\sx,\sa,h) |= \varphi$. 
	\item prove the following problem is decidable: given the formula $\varphi$ in PSeqSL, and a part of the model $\sx$, decide whether there exists $\sa,h$, such that $(\sx,\sa,h) |= \varphi$.
	\item prove the following problem is decidable: given the formula $\varphi$ in PSeqSL, decide whether there exists $\sx, \sa, h$, such that $(\sx,\sa,h) |= \varphi$. (\Cref{decidable_fragment_of_SeqSL})
\end{enumerate}

The first step is the main part of the proof. The result on the fragment without separating implications is trivial, because the finiteness of $h$ leads to finite possibilities of models satisfying the formula $\varphi$. 

Similar to the paper \cite{calcagno2001computability}, we define the size of the formula $\varphi$ as  $\sz(\varphi)$ to represent the maximum heap size for deciding whether the formula $\varphi$ is true or its negation is true. 

\begin{definition}[Size of formula]
	The size of the formula $\varphi$, $\sz(\varphi)$ can be inductively defined as follows: 
	\begin{table}[htbp]
		\centering  
		\label{table1}  
		\tabcolsep=0.5cm
		\renewcommand\arraystretch{1.2}
		\normalsize
		\begin{tabular}{ll}  
			$\sz(t_{x_1} = t_{x_2}) = 0$ &
			$\sz(t_{\alpha_1} == t_{\alpha_2}) = 0$ \\
			$\sz(\false) = 0$ &
			$\sz(\mathbf{emp}) = 1$ \\
			$\sz(x |-> t_\alpha) = 1$ &
			$\sz(\varphi_1 -* \varphi_2) = \sz(\varphi_2)$ \\
			\multicolumn{2}{l}{$\sz(\varphi_1 => \varphi_2) = \mathtt{max}(\sz(\varphi_1), \sz(\varphi_2))$} \\
			\multicolumn{2}{l}{$\sz(\varphi_1 * \varphi_2) = \sz(\varphi_1) + \sz(\varphi_2)$.}
		\end{tabular}
	\end{table}
\end{definition}

We define free program variables as follows.

\begin{definition}[Free program variables in PSeqSL]
	In PSeqSL, free program variables $\FV_{t,x}(t_x)$ in the term $t_x$ and free program variables $\FV_{t,x}(t_\alpha)$ in the term $t_\alpha$ can be inductively defined as follows:
	\begin{table}[htbp]
		\centering  
		\label{table1}  
		\tabcolsep=0.5cm
		\renewcommand\arraystretch{1.2}
		\small
		\begin{tabular}{lll}  
			$\FV_{t,x}(\nil) = \{\}$ &
			$\FV_{t,x}(\#) = \{\}$ &
			$\FV_{t,x}(x) = \{x\}$ \\
			$\FV_{t,x}(\alpha) = \{\}$ &
			$\FV_{t,x}(\varepsilon) = \{\}$ &
			\\
			\multicolumn{3}{l}{$\FV_{t,x}(t_{\alpha_{x_1}} \circ t_{\alpha_{x_2}}) = \FV_{t,x}(t_{\alpha_{x_1}}) \cup \FV_{t,x}(t_{\alpha_{x_2}})$.}
		\end{tabular}
	\vspace{-0.5em}
	\end{table}
	
	Free program variables $\FV_x(\varphi)$ in the formula $\varphi$ can be inductively defined as follows:
	\begin{align*}
		\FVl&(t_{x_1} = t_{x_2}) = 	\FV_{t,x}(t_{x_1}) \cup \FV_{t,x}(t_{x_2}) \\ 
		\FVl&(t_{\alpha_1} == t_{\alpha_2}) = \FV_{t,x}(t_{\alpha_1}) \cup \FV_{t,x}(t_{\alpha_2}) \\
		\FVl&(\false) = \{\} \\
		\FVl&(\varphi_1 => \varphi_2) = \FVl(\varphi_1) \cup \FVl(\varphi_2) \\
		\FVl&(\mathbf{emp}) = \{\} \\
		\FVl&(x |-> t_\alpha) = \{x\} \cup \FV_{t,x}(t_\alpha) \\
		\FVl&(\varphi_1 * \varphi_2) = \FVl(\varphi_1) \cup \FVl(\varphi_2) \\
		\FVl&(\varphi_1 -* \varphi_2) = \FVl(\varphi_1) \cup \FVl(\varphi_2).
	\end{align*}
\end{definition}

We define the set of sequence terms to collect the terms appearing in the formula. It helps us to get a finite set of all possible heap models which may satisfy the formula. 

\begin{definition}[Set of sequence terms]
	The set of sequence terms $\SeqTerms(\varphi)$ appearing in the formula $\varphi$ can be inductively defined as:
	\begin{align*}
		\SeqTerms&(t_{l_1} = t_{l_2}) = \{t_{l_1}, t_{l_2}\} \\
		\SeqTerms&(t_{\alpha_1} == t_{\alpha_2}) = \{t_{\alpha_1}, t_{\alpha_2}\} \\
		\SeqTerms&(\false) = \{\} \\
		\SeqTerms&(\mathbf{emp}) = \{\} \\
		\SeqTerms&(x |-> t_\alpha) = \{x, t_\alpha\} && \\
		\SeqTerms&(\varphi_1 => \varphi_2) = \SeqTerms(\varphi_1) \cup \SeqTerms(\varphi_2) && \\
		\SeqTerms&(\varphi_1 * \varphi_2) = \SeqTerms(\varphi_1) \cup \SeqTerms(\varphi_2) \\ \SeqTerms&(\varphi_1 -* \varphi_2) = \SeqTerms(\varphi_1) \cup \SeqTerms(\varphi_2).
	\end{align*}
\end{definition}

Separating implication involves universal quantifiers over heaps. The paper \cite{calcagno2001computability} presents a solution on searching all possible models of the formula in the classical $\Sigma_1$ separation logic fragment. They restrict the domain and the range of the heap $h$ to finite sets. To deal with sequences, we come up with a different restriction on the range of $h$. The range of $h$ involves the set of all sequence terms appearing in the formula, empty sequence $\varepsilon$, and a fresh sequence term whose value is different from all sequence terms in the formula. Thus, universal quantifiers over heaps can be reduced to those over finite heaps. The reduction preserves satisfiability. 

We present the following lemma to show small model property on the formula $\varphi_1 -* \varphi_2$.

\begin{lemma}\label{small_model_property_of_si}
	Given the model $\sigma = (\sx,\sa,h)$ and formulae $\varphi, \; \varphi_1 \text{ and } \varphi_2$ satisfying $\varphi = \varphi_1 -* \varphi_2$, the set $L = \FVl(\varphi_1) \cup \FVl(\varphi_2)$, the set $B$ which contains the first $\mathtt{max}(\sz(\varphi_1), \sz(\varphi_2))$ values in $\mathrm{Loc} \setminus (\dom(h) \cup s(L))$, the fresh sequence variable $\overline{\beta}$ satisfying $|[\overline{\beta}|]\sigma' \notin \{|[t_\alpha|]\sigma' \mid t_\alpha \in \mathtt{SeqTerms}(\alpha)\}$. We define $\sigma' = (\sx,\sa',h)$ where $\sa'$ contains a new assignment for $\overline{\beta}$. Let $\mathcal{D} = B \cup s(L)$, and $\mathcal{R} = \{|[t_\alpha|]\sigma' \mid t_\alpha \in \SeqTerms(\varphi)\} \cup \{\varepsilon, |[\overline{\beta}|]\sigma'\}$. Then, $(\sx,\sa,h) |= \varphi_1 -* \varphi_2$ holds, if and only if for all $h'$ satisfying:
	\begin{itemize}
		\item $\dom(h') \cap \dom(h) = \varnothing$ and $(\sx,\sa',h') |= \varphi_1$,
		\item $\dom(h') \subseteq \mathcal{D}$,
		\item $\mathtt{rng}(h') \subseteq \mathcal{R}$, 
	\end{itemize}
	the proposition $(\sx,\sa',h \uplus h') |= \varphi_2$ holds, where $\mathtt{rng}(h')$ denotes the range of $h'$. 
\end{lemma}


According to \Cref{small_model_property_of_si}, we define the reduction function as follows. 
\begin{definition}\label{reduction_function}
	The reduction function $T(\sx,h,\varphi)$ is a mapping from stack $\sx$, heap $h$, and the PSeqSL formula $\varphi$, to the formula in $\Sigma_1$ sequence predicate logic. The reduction can be inductively defined in Fig. (\ref{figeq:redfunc}),
	\stripsep+2pt
	\begin{figure*}[h]
	\begin{align*}
		T(\sx,h,t_{x_1} = t_{x_2}) &\quad\overset{\triangle}{=}\quad t_{x_1} == t_{x_2} \\
		T(\sx,h,t_{\alpha_1} == t_{\alpha_2}) &\quad\overset{\triangle}{=}\quad t_{\alpha_1} == t_{\alpha_2} \\
		T(\sx,h,\false) &\quad\overset{\triangle}{=}\quad \false \\
		T(\sx,h,\varphi_1 => \varphi_2) &\quad\overset{\triangle}{=}\quad T(\sx,h,\varphi_1) => T(\sx,h,\varphi_2) \\
		T(\sx,h,\mathbf{emp}) &\quad\overset{\triangle}{=}\quad \dom(h) = \varnothing \\
		T(\sx,h,x|->t_\alpha) &\quad\overset{\triangle}{=}\quad \dom(h)=\{\sx(x)\} \land \sx(x) \notin \mathrm{Atom} \land h(\sx(x)) == t_\alpha \\
		T(\sx,h,\varphi_1 * \varphi_2) &\quad\overset{\triangle}{=}\quad \bigvee_{h = h_1 \uplus h_2} (T(\sx,h_1,\varphi_1) \land T(\sx,h_2,\varphi_2)) \\
		T(\sx,h,\varphi_1 -* \varphi_2) &\quad\overset{\triangle}{=}\quad \bigwedge_{
			\mbox{ \tiny
				$\begin{array}{c}
					\dom(h_\varphi) \cap \dom(h) = \varnothing\\
					\dom(h_\varphi) \subseteq \mathcal{D} \\
					\mathtt{rng}(h_\varphi) \subseteq \mathcal{R}
				\end{array}
				$
			}
		} (T(\sx,h_\varphi,\varphi_1) => T(\sx,h_\varphi \uplus h,\varphi_2)), 
	\end{align*}
	\captionof{figure}{The reduction function $T(\sx,h,\varphi)$}
	\label{figeq:redfunc}
	\end{figure*}
	where $\varphi = \varphi_1 -* \varphi_2$. 
\end{definition} 

Given $\sx$ and $h$, the formula $T(\sx,h,\varphi_1 * \varphi_2)$ consists of finitely many terms of the form $T(\sx,h,\varphi)$, because the implicit existential quantifier in separating conjunction does not create new heaps. The formula $T(\sx,h,\varphi_1 -* \varphi_2)$ also consists of finitely many terms of the form $T(\sx,h,\varphi)$ according to \Cref{small_model_property_of_si}. Moreover, the truth values of the following three formulae $\dom(h) = \varnothing$, $\dom(h) = {\sx(x)}$, and $\sx(x) \notin \mathrm{Atom}$ can be determined, as well as the value of $h(\sx(x))$ (which is a sequence). So $T(\sx,h,\varphi)$ consists of only $t_{\alpha_1} == t_{\alpha_2}$ and its conjunctions, disjunctions and negations. It can be further reduced to a single word equation followed by \Cref{thm:Boolean comb}. PSeqSL is thus reduced to the satisfiability problem of $\Sigma_1$ sequence predicate logic. The $\Sigma_1$ fragment of sequence predicate logic is shown to be decidable in \cite{makanin1977problem}. 

\begin{lemma}\label{lemma:part1}
	Given stack $\sx$ and heap $h$. For all assignment $\sa$ and formula $\varphi$, the proposition $(\sx,\sa,h) |= \varphi$ holds if and only if there exist assignments $(\sx', \sa')$, such that $(\sx', \sa') |= T(\sx,h,\varphi)$ holds, where $\sx'$ and $\sa'$ are assignments of program variables and sequence variables respectively, $T(\sx,h,\varphi)$ is the reduction from  PSeqSL to $\Sigma_1$ sequence predicate logic defined by \Cref{reduction_function}.
\end{lemma}

Note that after doing reductions in \Cref{reduction_function} and applying \Cref{thm:Boolean comb}, one can get a single word equation $U == V$ with constants ($k$ contents for example) and sequence variables ($k'$ variables for example). It is easy to show the following fact: there is a solution for $U == V$ if and only if there is a solution for $U == V$ on the free monoid with $k+k'$ generators, where the first $k$ generators coincide with $k$ constants, and the other $k'$ generators are fresh in $\mathbb{N}$.

To give readers some intuitions for \Cref{lemma:part1}, we list two examples below.
\begin{example}
	Consider whether the formula $\varphi = (x_1 |-> \alpha_1 \circ x_3) * \big(
	(x_1 |-> \alpha_1 \lor x_1 |-> \alpha_2) -*
	(x_1 |-> \alpha_2 * x_2 |-> \alpha_3)
	\big)$ can be satisfied, given the stack satisfying $\{(x_1,n_1),(x_2,n_2),(x_3,n_3)\} \subseteq s$ where $n_1,n_2$ and $n_3$ are distinct integer numbers and the heap $h = \{(n_1,n_1 \circ n_3), (n_2,n_2 \circ n_3)\}$. The assignment $\sa$ is pending.
	
	Suppose $\varphi = \varphi_1 * \varphi_2$, where $\varphi_1 = x_1 |-> \alpha_1 \circ x_3$, $\varphi_2 = 
	(x_1 |-> \alpha_1 \lor x_1 |-> \alpha_2) -*
	(x_1 |-> \alpha_2 * x_2 |-> \alpha_3)$. We have the formula in Fig. (\ref{figeq:cons_ex1}). 

\stripsep+0.5em
\begin{figure*}[h]
	\begin{align*}
		T(\sx,h,\varphi) \quad=\quad& \bigvee_{h = h_1 \uplus h_2}(T(\sx,h_1,\varphi_1) \land T(\sx,h_2,\varphi_2)) \\
		\quad=\quad& \; \big( T(\sx,\{\},\varphi_1) \land T(\sx,\{(n_1,n_1 \circ n_3), (n_2,n_2 \circ n_3)\},\varphi_2) \big) \\
		&  \lor \big( T(\sx,\{(n_1,n_1 \circ n_3)\},\varphi_1) \land T(\sx,\{(n_2,n_2 \circ n_3)\},\varphi_2) \big) \\
		&  \lor \big( T(\sx,\{(n_2,n_2 \circ n_3)\},\varphi_1) \land T(\sx,\{(n_1,n_1 \circ n_3)\},\varphi_2) \big) \\
		&  \lor \big( T(\sx,\{(n_1,n_1 \circ n_3),(n_2,n_2 \circ n_3)\},\varphi_1) \land T(\sx,\{\},\varphi_2) \big) \\
		\quad=\quad& \; T(\sx,\{(n_1,n_1 \circ n_3)\},\varphi_1) \land T(\sx,\{(n_2,n_2 \circ n_3)\},\varphi_2).
	\end{align*}
	\caption{Construction in Example 1}
	\label{figeq:cons_ex1}
\end{figure*}
	Suppose $\varphi_2 = \varphi_{21} -* \varphi_{22}$, where $\varphi_{21} = x_1 |-> \alpha_1 \lor x_1 |-> \alpha_2$, and $\varphi_{22} = x_1 |-> \alpha_2 * x_2 |-> \alpha_3$. We have:
	\vspace{-1em}
	\begin{equation*}
		\begin{aligned}
			& T(\sx,\{(n_2,n_2 \circ n_3)\},\varphi_2) \\
			=\quad &\bigwedge_{
				\mbox{ \tiny
					$\begin{array}{c}
						\dom(h_{\varphi_2}) \cap \{n_2\} = \varnothing\\
						\dom(h_{\varphi_2}) \subseteq \mathcal{D} \\
						\mathtt{rng}(h_{\varphi_2}) \subseteq \mathcal{R}
					\end{array}
					$
				}
			} \bigl(T(\sx,h_{\varphi_2},\varphi_{21}) \\
			& => T(\sx,h_{\varphi_2} \uplus \{(n_2,n_2 \circ n_3)\},\varphi_{22})\bigr), 
		\end{aligned}
	\vspace{-1em}
	\end{equation*}
	where
	\vspace{-0.5em}
	\begin{equation*}
	\begin{aligned}
		\mathcal{D} &= B \;\cup\; \{n_1,n_2,n_3\}, \quad B = \{m_1,m_2,m_3\}, \\
		\mathcal{R} &= \{|[\alpha_1 \circ x_3|]\sigma', |[\alpha_1|]\sigma',|[\alpha_2|]\sigma', |[\alpha_3|]\sigma'\} \cup \{\varepsilon, |[\overline{\beta}|]\sigma'\},
	\end{aligned}
	\vspace{-1em}
	\end{equation*}

	and $m_1,m_2,m_3$ are the first three values in $\mathrm{Loc} \setminus (\dom(h) \cup s(L))$, the term $\overline{\beta}$ satisfies:
	\begin{equation*}
	\begin{aligned}
		|[\overline{\beta}|]\sigma' \notin \{|[\alpha_1 \circ x_3|]\sigma', |[\alpha_1|]\sigma',|[\alpha_2|]\sigma', |[\alpha_3|]\sigma'\}.
	\end{aligned}
	\end{equation*}

	The domain $\dom(h_{\varphi_2})$ consists of many finite possibilities. However, there are only two possibilities $h_{\varphi_2} = \{(n_1,\sa'(\alpha_1))\}$ and $h_{\varphi_2} = \{(n_1,\sa'(\alpha_2))\}$ which may satisfy the formula. So, 
	\begin{equation*}
	\begin{aligned}
		& T(\sx,\{(n_2,n_2 \circ n_3)\},\varphi_2) \\
		=\quad& \; (T(\sx,\{(n_1,\sa'(\alpha_1)\},\varphi_{21}) \\
		& => T(\sx,\{(n_1,\sa'(\alpha_1)), (n_2,n_2 \circ n_3)\},\varphi_{22})) \\
		& \land (T(\sx,\{(n_1,n_2 \circ n_3)\},\varphi_{21}) \\
		& => T(\sx,\{(n_1,n_2 \circ n_3), (n_2,n_2 \circ n_3)\},\varphi_{22})) \\
		=\quad& \; T(\sx,\{(n_1,\sa'(\alpha_1)), (n_2,n_2 \circ n_3)\},\varphi_{22}) \\
		& \land T(\sx,\{(n_1,n_2 \circ n_3), (n_2,n_2 \circ n_3)\},\varphi_{22}).
	\end{aligned}
	\end{equation*}
	
	Similarly, we discuss possibilities of models satisfying the formula $\varphi_{22}$. Then we have:
	\begin{align*}
		&T(\sx,\{(n_1,\sa'(\alpha_1)), (n_2,n_2 \circ n_3)\},\varphi_{22}) \\
		=\quad& \alpha_1 == \alpha_2 \land n_2 \circ n_3 == \alpha_3,\\
		&T(\sx,\{(n_1,n_2 \circ n_3), (n_2,n_2 \circ n_3)\},\varphi_{22}) \\
		= \quad& \alpha_2 == \alpha_2 \land n_2 \circ n_3 == \alpha_3.
	\end{align*}
	
	So, 
	\begin{align*}
		&T(\sx,h,\varphi) \\
		=\quad& (n_1 \circ n_3 == \alpha_1 \circ n_3) \land (\alpha_1 == \alpha_2 \land n_2 \circ n_3 == \alpha_3) \\
		& \land (\alpha_2 == \alpha_2 \land n_2 \circ n_3 == \alpha_3).
	\end{align*}
	
	We observe that $T(\sx,h,\varphi)$ can be satisfied. Hence given $\sx,h$, there exists $\sa$ such that $(\sx,\sa,h) |= \varphi$ holds.
\end{example}

\begin{example}
	Consider  whether the formula $\varphi = x_1 |-> \alpha_1 \circ x_3 \land \big( 
	(x_2 |-> \alpha_1 \lor x_2 |-> \alpha_2) -* 
	(x_1 |-> \alpha_1 * x_2 |-> \alpha_3)
	\big)$ can be satisfied, given $s = \{(x_1,n_1), (x_2,n_2), (x_3,n_3)\}, h = \{(x_1,n_1 \circ n_3)\}$. 
	
	Following definition Definition 4.4, we have:
	\begin{align*}
		&T(\sx,h,\varphi) \\
		=\quad& n_1 \circ n_3 == \alpha_1 \circ n_3 \land \big(
		(\alpha_1 == \alpha_3 \land \alpha_1 == n_1 \circ n_3) \\
		& \land (\alpha_2 == \alpha_3 \land \alpha_1 == n_1 \circ n_3)
		\big).
	\end{align*}
	We observe that $T(\sx,h,\varphi)$ cannot be satisified. Hence given $\sx,h$, there does not exist $\sa$ such that $(\sx,\sa,h) |= \varphi$ holds. We omit the details here. 
\end{example}

For the second step of the proof, heap $h$ is not given beforehand. The satisfiability problem of the formula in PSeqSL can be reduced to the problem in the first step by the following lemma.

\begin{lemma}\label{cor:sat}
	Given stack $\sx$ and the PSeqSL formula $\varphi$, the following problem is decidable: whether there exists $\sa$ and $h$ such that $(\sx,\sa,h) |= \varphi$.
\end{lemma}

The above problem is equivalent to the following problem: given $\sx$, whether there exists $\sa$ such that $(\sx,\sa,\{\}) |= \varphi \septra \true$ holds. Thus we can conclude \Cref{cor:sat}. 

For the third step, stack $\sx$ is not given. Similar to the paper\cite{calcagno2001computability}, we need to figure out a finite range of $\sx$, such that the formula is true on the range if and only if the formula is true on infinite range of $\sx$. 

\begin{definition}
	Given two models $(\sx,\sa,h), (\sx',\sa',h')$, the set $P \subseteq \mathrm{PVar}$, and the set $S \subseteq \mathrm{SVar}$. The relation $(\sx,\sa,h) \approx_P (\sx',\sa',h')$ holds, if and only if there exists an isomorphism $r:\; (\mathrm{Loc}^*, \circ) -> (\mathrm{Loc}^*, \circ)$, such that the following conditions are satisfied:
	\begin{itemize}
		\item for each $\alpha_1,\alpha_2 \in S$, we have $r(\sa(\alpha_1) \circ \sa(\alpha_2)) = r(\sa(\alpha_1)) \circ r(\sa(\alpha_2))$,
		\item $r(\nil) = \nil$, $r(\#) = \#$, and for each $x \in P$, we have $r(\sx(x)) = \sx'(x)$,
		\item for each $x \in P$, we have $r(h(x)) = h'(r(x))$,
		\item for each $\alpha \in S $, we have $r(\sa(\alpha)) = \sa'(\alpha)$.
	\end{itemize}
\end{definition}

\begin{proposition}
	Given two models $(\sx,\sa,h), (\sx',\sa',h')$, and the formula $\varphi$ satisfying $(\sx,\sa,h) \approx_{\FVl(\varphi)} (\sx',\sa',h')$. If $(\sx,\sa,h) |= \varphi$, then $(\sx',\sa',h') |= \varphi$.
\end{proposition}

\begin{lemma}\label{lemma:part3}
	Given the model $(\sx,\sa,h)$ and the formula $\varphi$. Let $B$ be the first $|\FVl(\varphi)|$ locations in $\mathrm{Loc}$. There exists $(\sx',\sa',h')$ such that $\sx'(\mathrm{PVar} \setminus \FVl(\varphi)) \subseteq \{\nil,\#\}, \sx'(\FVl(\varphi)) \subseteq B \cup \{\nil,\#\}$, and $(\sx,\sa,h) \approx_{\FVl(\varphi)} (\sx',\sa',h')$. 
\end{lemma}

With \Cref{lemma:part1,cor:sat,lemma:part3}, we can conclude that the satisfiability problem of PSeqSL is decidable, that is \Cref{theorem:decidable_fragment} holds. 

As we know, the truth of $\Sigma_1$ fragment of sequence predicate logic is decidable\cite{makanin1977problem}. As the fragment consists of negations without any restrictions, there is a one-to-one mapping from instances in $\Sigma_1$ fragment of sequence predicate logic and those in $\Pi_1$ fragment of sequence predicate logic. Thus we have \Cref{cor:seq}.
\begin{corollary}\label{cor:seq}
	The theory of $\Pi_1$ fragment of sequence predicate logic is decidable. 
\end{corollary}

The proof of \Cref{theorem:decidable_fragment} shows that PSeqSL has small model property. The satisfiability of a formula in PseqSL can be reduced to that of the formula with finite length in $\Sigma_1$ fragment sequence logic. According to small model property of PSeqSL and \Cref{cor:seq}, we have \Cref{cor:p1_frag}.
\begin{corollary}\label{cor:p1_frag}
	The satisfiability and validity problem of $\Pi_1$ fragment of PSeqSL is decidable, where $\Pi_1$ fragment of PSeqSL is of the form $\forallx^*\foralla^*.\varphi$, and $\varphi$ is quantifier-free. 
\end{corollary}


\section{Undecidable fragments}

In this section, we mainly focus on an undecidable fragment of the form
$(\forallx^*\foralla^*\cap \forallx^*\existsx^*\existsa^*)\text{SeqSL}(*)$. It is the conjunction of formula in $(\forallx^* \foralla^*)\text{SeqSL}(*)$ and formula $(\forallx^*\existsx^*\existsa^*)\text{SeqSL}(*)$, where $\existsa^*$ denotes there are 0 or more existential quantifiers over sequence variables, and $\existsx^*$ denotes there are 0 or more existential quantifiers over program variables. The meaning of $\forall_\alpha^*$ and $\forall_\alpha^*$ are similar. We present the following main result of this section. 

\begin{theorem}[Undecidable fragment of SeqSL]\label{thm:und}
	The satisfiability problem of the language is undecidable when it is restricted as follows:
	\begin{align*}		
		t_x \quad::=\quad& \nil \mid \# \mid x \\
		t_\alpha \quad::=\quad& \varepsilon \mid t_x \mid \alpha \mid t_\alpha \circ t_\alpha \\
		\psi \quad::=\quad& t_x = t_x \mid t_\alpha == t_\alpha \mid x \hookrightarrow t_\alpha \mid \false \mid \psi => \psi \\
		& \mid \emp \mid \psi * \psi \\
		\varphi \quad::=\quad& \forallx^* \foralla^*. \psi \land \forallx^*\existsx^*\existsa^*. \psi,
	\end{align*}
	where $\varphi$ is a sentence.
	
	The model of the fragment is defined in \Cref{def:modelSeqSL}.
\end{theorem}

We reduce the problem from halting problem of two-counter Minsky machine. Before getting into details of the reduction, we recall the definition of two-counter Minsky machine. 
\begin{definition}[Two-counter Minsky machine]\label{def_Minsky}
	Let $M$ be a Minsky machine with $n >= 1$ instructions. The machine $M$ has two counters $C_1$ and $C_2$. The instructions are defined as follows:
	\begin{enumerate}[]
		\item $I:\; C_j := C_j + 1;$ goto $k$,
		\item $I:$ if $C_j = 0$, then goto $k_1$, else ($C_j := C_j - 1$; goto $k_2$),
		\item $n:$ halt,
	\end{enumerate}
	where $j \in [1,2],\; I \in [1, n-1],$ and $k,k_1,k_2 \in [1, n]$. Machine $M$ halts if there is a run of the form
	\begin{align*}
		(I_0, c_0^1, c_0^2),\; (I_1, c_1^1, c_1^2),\dots, (I_m, c_m^1, c_m^2), 
	\end{align*}
	such that $(I_i, c_i^1, c_i^2) \in [1,n] \times \mathbb{N}^2\; (i \in [1,m])$, $I_0 = 1,\; I_m = n$, and $c_0^1 = c_0^2 = 0$. The run follows the instructions defined above.
	
	The set of instructions with type 1 (denoted by $\mathcal{I}_1$) consists of tuples $(k_0,c_j,c_j',k)$, and those with type 2 (denoted by $\mathcal{I}_2$) consists of tuples $(k_0,c_j,c_j',k_1,k_2)$, where $k_0$ denotes the current pointer, $c_j,c_j' \in [1,2]$ denote two counters ($c_j != c_j'$), and $k,k_1,k_2$ denote the next pointers. 
\end{definition}

The problem of deciding whether a machine $M$ halts is known to be undecidable\cite{minsky1968computation}. 

We first show the undecidable fragment defined in \Cref{thm:und} is not that easy to get. In general, decidability results of both sequence logic fragments and separation logic fragments restrict the form of undecidable fragments of SeqSL. 

Decidability results in free semigroup show that, the theory of $\Sigma_2$ fragment\cite{day2018satisfiability}, the positive theory of $\Pi_2$ fragment\cite{durnev1995undecidability} and $\Sigma_3$ fragment\cite{durnev1995undecidability} of word equations are undecidable. The results in separation logic show that, the satisfiability problem of $\Sigma_3$ fragment of separation logic is undecidable. In this case, We can get \Cref{fact:trival_und_fragments}. 

\begin{fact}\label{fact:trival_und_fragments}
	In the fragment of SeqSL with prenex normal form, if there are 3 or more alternations of quantifiers over program variables, or 2 or more over sequence variables, then the satisfiability problems of resulting fragments are undecidable. For instance, the satisfiability problems of  $(\existsx^*\existsa^*\foralla^*)\mathrm{SeqSL}$, $(\existsx^*\foralla^*\existsx^*)\mathrm{SeqSL}$, $(\existsx^*\forallx^*\foralla^*\existsx^*)\mathrm{SeqSL}$ are undecidable. 
\end{fact}

In order to get a non-trivial undecidable fragment of SeqSL with negations, we have to restrict the alternations of quantifiers over program variables to be less than 3, and those over sequence variables to be less than 2. 

\begin{figure}[h]
		\centering
		\includegraphics[width=0.45\textwidth]{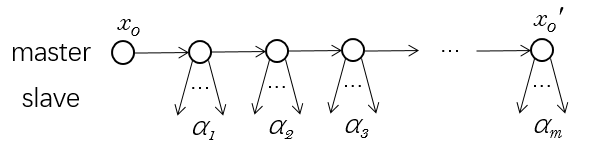}
		\captionof{figure}{Structure of sapling}
		\label{fig:sapling}
	\end{figure}

To prove \Cref{thm:und}, we first discuss the encoding of the reduction. The general structure is sapling, which is shown in Fig. (\ref{fig:sapling}).

We consider the encoding for sapling. Sapling is a fishbone-like structure with 'bones' going in opposite directions. Sapling consists of a master branch and several slave branches attached to nodes in the master branch. The depth of each slave branch is 1, and the end points of them does not point to any nodes in master branch. Similar to the encoding for the  list $x \overset{\circlearrowleft}{\longrightarrow}^+ y$ defined in \cite{brochenin2012almighty} and  fishbone heap defined in \cite{demri2015two}, we encode the sapling in the following ways:
\begin{enumerate}
	\item There are smaller or equal than 1 predecessor of each node. It is denoted by the formula $\varPsi_1$,
	\item There is no predecessor on the first master node. It is denoted by the formula $\varPsi_2$,
	\item The first master node is allocated, and the last node does not point to the next master node. It is denoted by the formula $\varPsi_3$ (it is not necessary to make the predecessor of the last node to be 1),
	\item For each node except for the last node, there is another allocated node which follows the node. It is denoted by the formula $\varPsi_4$. 
\end{enumerate}

The sapling from node $x_0$ to node $x_0'$ (assume here that $x_0 != x_0'$) can be expressed by 
\begin{align}\label{eq:sap}
	\mathtt{sapling}(x_0,x_0') &\overset{\triangle}{=}  \varPsi_1 \land \varPsi_2 \land \varPsi_3 \land \varPsi_4. 
\end{align}
where $\varPsi_1, \varPsi_2, \varPsi_3$, and $\varPsi_4$ are defined in Fig. (\ref{figeq:cons_ex1}).

\stripsep+0.5em
\begin{figure*}[h]
\begin{equation*}
	\begin{aligned}
		\varPsi_1 &\quad\overset{\triangle}{=}\quad \forallx x_1\forallx x_2\forallx x_3\forallx x_4 \foralla \alpha_1\foralla \alpha_2.\; (x_1 \hookrightarrow x_3 \circ \alpha_1 * x_2 \hookrightarrow x_4 \circ \alpha_2 => x_3 != x_4) \\
		\varPsi_2 &\quad\overset{\triangle}{=}\quad \forallx x_1\foralla \alpha.\; \lnot (x_1 \hookrightarrow x_0 \circ \alpha) \\
		\varPsi_3 &\quad\overset{\triangle}{=}\quad (\existsx x_1.\; x_0 \hookrightarrow x_1 \circ \varepsilon) \land (\existsa \alpha.\; x_0' \hookrightarrow \varepsilon \circ \alpha) \\
		\varPsi_4 &\quad\overset{\triangle}{=}\quad \forallx x_1\forallx x_2\existsx x_3\existsa \alpha_1\existsa \alpha_2.\; \bigl((x_1 \hookrightarrow x_2 \circ \alpha_1 \land x_2 != x_0') => x_2 \hookrightarrow x_3 \circ \alpha_2\bigr). 
	\end{aligned}
\end{equation*}
\caption{Definition of $\varPsi_1, \varPsi_2, \varPsi_3$, and $\varPsi_4$}
\label{key}
\end{figure*}

It can be easily shown that the predicate $\mathtt{sapling}(x_0,x_0')$ is of the form $(\forallx^* \foralla^* \cap \forallx^* \existsx^* \existsa^*)\varPsi$. 

Note that the encoding in \Cref{eq:sap} is just a general idea. It does not prevent contents in slave nodes from pointing to other nodes. However, this does not have impacts on the main result, because these contents only consist of $\nil$ which will be shown in \Cref{formula_und_1}. 

Note also that the predicate $\mathtt{Sapling}$ not only encodes the graph consisting of only one sapling, but also graphs consisting of both one sapling and circles. It is not necessary to remove all these circles, because it does not have impacts on correctness of the reduction. It is shown in \Cref{lemma:sap}.
\begin{lemma}\label{lemma:sap}
	Given the predicate $\mathtt{sapling}(x_0,x_0')$ defined in \Cref{eq:sap}, it can be satisfied by a model $\sigma$ if and only if there exists a model $\sigma'$ that satisfies $\mathtt{sapling}(x_0,x_0')$, and there are no circles in $\sigma'$.
\end{lemma}

The proof of \Cref{lemma:sap} can be found in Appendix. Now we prove \Cref{thm:und}. 

%
%

\begin{proof}[of \Cref{thm:und}]
We go into details of the reduction, which is shown in Fig. (\ref{fig:reduction}). 
\begin{figure}[h]
	\centering
	\includegraphics[width=0.5\textwidth]{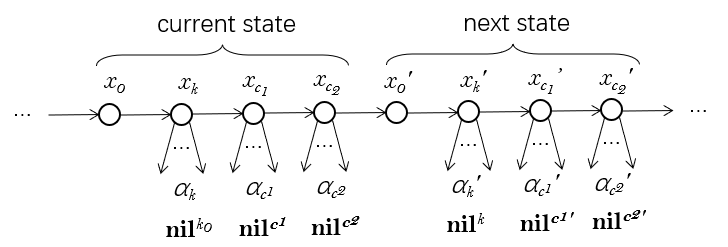}
	\caption{Structure of reduction}
	\label{fig:reduction}
\end{figure}

The master branch is the longest path from the beginning master node to the end master node. Each master node points to 0 or more slave nodes $\nil^*$. The master nodes have the period of 4. Each period represents a state of the run. In the $i$-th period ($1 <= i <= m$), the first master node represents a deliminator which separates each state of the run. There is no slave node attached to it. The second to the fourth master nodes represent the $i$-th triple $(I,C_1,C_2)$ satisfying $I = k_i,\; C_1 = c_i^1 +1, \text{ and } C_2 = c_i^2 + 1$. The slave nodes pointed from these 3 nodes are $\nil^{k_i},\;  \nil^{c_i^1 + 1}, \text{ and } \nil^{c_i^2 + 1}$ respectively. The first period represents the initial state, while the last represents the final state.

We observe that the values of two counters in the sapling are greater than the real values of the counters in the run by 1. It is because we need to distinguish these nodes from the first node of each period, to which there is no slave nodes attached. 

In general, The reduction formula consists of the following three parts. 
\begin{enumerate}
	\item Construct the basic structure of sapling, which is denoted by the formula $\varPhi_1 = \forallx^*\foralla^*.\varPhi_1'$.
	\item Initialize the initial and the final state, which is denoted by the formula $\varPhi_2 = \existsx^*\existsa^*.\varPhi_2'$.
	\item Encode the transition from one state to the other following the current instruction, which is denoted by the formula $\varPhi_3 = \forallx^*\existsx^*\existsa^*.\varPhi_3'$. 
\end{enumerate}
The formula is of the form
\begin{equation}\label{formula_und_1}
\begin{aligned}
	\varPhi \quad=\quad& \varPhi_1 \land \varPhi_2 \land \varPhi_3 \\
	\quad=\quad& \forallx^*\foralla^*.\varPhi_1' \land \existsx^*\existsa^*.\varPhi_2' \land \forallx^*\existsx^*\existsa^*.\varPhi_3', 
\end{aligned}
\end{equation}
or
\begin{align*}
	\varPhi \quad=\quad \forallx^*\foralla^*.\varPhi_1' \land \forallx^*\existsx^*\existsa^*.(\varPhi_2' \land \varPhi_3'). 
\end{align*}

The detailed construction can be found in Appendix.

For each two-counter Minsky machine M, let $\varPhi$ be the formula of the form defined in \Cref{formula_und_1}. If the machine M halts, then there is a finite run. It can be easily shown that there is a finite sapling from the first period encoding the first state to the final one encoding the final state, and each state except for the final state can be transformed to the next state by the corresponding instruction. 
	
For the other side of the proof, we suppose that the formula $\varPhi$ of the form defined in \Cref{formula_und_1} is satisfied. The model $\sigma$ of $\varPhi$ may consist of circles. We can get a model $\sigma'$ satisfying $\varPhi$ in which there is no circles according to \Cref{lemma:sap}. 
\end{proof}

\section{Conclusion}
In this paper, we propose sequence-heap separation logic which combines sequence predicate logic and separation logic. It is capable of describing the following properties which sequence predicate logic or separation logic alone cannot describe or cannot easily describe. 
\begin{itemize}
	\item sequence operations in programs, such as list reversal, and lookups.
	\item properties corresponding to variable-length sequences in programs, such as stack, queue, and graphs with unbounded out-degree.
	\item multilevel data structures in programs, such as data structure in Windows storage systems, and in block-based cloud storage systems.
\end{itemize}

Besides, we find a boundary between decidable and undecidable SeqSL fragments, which are both of the prenex normal form. We prove the following two decidable results: the satisfiability problem of $\Sigma_1$ fragment and $\Pi_1$ fragment is decidable, and that of $(\forallx^*\foralla^*\cap \forallx^*\existsx^*\existsa^*)\text{SeqSL}(*)$ is undecidable. As corollaries,  fragments with either 3 quantifier alternations over program variables or 2 quantifier alternations over sequence variables are undecidable. 

We will do the following work in the future.
\begin{itemize}
	\item Find other boundaries between decidable and undecidable fragments of sequence-heap separation logic, such as boundaries on numbers of quantified variables, and on inductive predicates.
	\item Investigate deeply on expressiveness of sequence-heap separation logic.
	\item Construct a proof system for sequence-heap separation logic.
	\item Implement formal verification tools for sequence-heap separation logic fragments.
\end{itemize}

\bibliographystyle{IEEEtran} 
\bibliography{bibl.bib}

\onecolumn
\begin{center}
	\Large
	Appendix
\end{center}
\vspace{3em}
\textit{Proof sketch of \Cref{thm:Boolean comb}:}
	For every Boolean combinations $\varphi$ defined in \Cref{def:Boolcomb}, there exists a single word equation $T(\varphi) \overset{\triangle}{=} U == V$, which is equivalent to $\varphi$ in the following inductive way:
	\begin{align*}
		T&(t_1 == t_2) &\overset{\triangle}{=}\quad& t_1 == t_2 \\
		T&(t_1 == t_2 \land t_1' == t_2') &\overset{\triangle}{=}\quad& F(t_1,t_2) == F(t_1',t_2') \\
		T&(t_1 == t_2 \lor t_1' == t_2') &\overset{\triangle}{=}\quad& T_0(t_1 \circ t_2' == t_2 \circ t_2' \lor t_2 \circ t_1' == t_2 \circ t_2') \\
		T&(\lnot\; t_1 == t_2) &\overset{\triangle}{=}\quad& \exists \beta\exists \beta'\exists \beta''.(\bigvee_{n \in \Sigma}t_1 == t_2 \circ n \circ \beta) \lor (\bigvee_{n \in \Sigma}t_2 == t_1 \circ n \circ \beta) \\
		& & & \lor (\bigvee_{\mbox{\scriptsize
					$\begin{array}{c}
						n,n' \in \Sigma \\
						n!=n'
					\end{array}
					$
			}
		} t_1 == \beta \circ n \circ \beta' \land t_2 == \beta \circ n' \circ \beta'') \\
		T&_0(u == v \lor u' == v) &\overset{\triangle}{=}\quad& \exists \beta \exists \beta'.X == \beta \circ Y \circ \beta',
	\end{align*}
	where $\Sigma$ is generator set, and
	\begin{align*}
		X &\quad=\quad G(u \circ u')^2 \circ u \circ G(u \circ u')^2 \circ u' \circ G(u \circ u')^2 &\\
		Y &\quad=\quad G(u \circ u')^2 \circ v \circ G(u \circ u')^2 &\\
		F(t_1, t_2) &\quad=\quad t_1 \circ n \circ t_2 \circ t_1 \circ n' \circ t_2,\;& n != n' \\
		G(t) &\quad=\quad t \circ n \circ t \circ n',\; &n != n'
	\end{align*}
	$\hfill \square$
\\
\\
\textit{Proof sketch of \Cref{lemma:sap}:} 
	The 'if' part of the lemma is straightforward. For the 'only if' part, we suppose that $\sigma |= \mathtt{sapling}(x_0,x_0')$. It is sufficient to show the following two facts.
	\begin{itemize}
		\item It is impossible that all connected components in $\sigma$ are circles,
		\item If there are one or more circles in $\sigma$, then let $\sigma'$ be the model with only one sapling in $\sigma$ and without circles. We can get $\sigma' |= \mathtt{sapling}(x_0,x_0')$. 
	\end{itemize}
	
	First we prove the first fact. We suppose that all connected components in $\sigma$ are circles. We can imply that each master node has one predecessor. It contradicts $\varPsi_2$ which claims there are no predecessors on the first node $x_0$.
	
	Then we prove the second fact. According to the first fact, there exists one sapling in $\sigma$. The sapling does not intersect with other circles according to $\varPsi_1$ which claims that each master node has no more than 2 predecessors. Let $\sigma'$ be the sapling in $\sigma$, and $\sigma''$ be all circles in $\sigma$. We can have $\sigma = \sigma' \uplus \sigma''$. So, the proposition $\sigma' |= \mathtt{sapling}(x_0,x_0')$ is true, because removing circles does not affect the truth of $\varPsi_1,\varPsi_2,\varPsi_3$, and $\varPsi_4$.
\hfill $\square$
\\
\\
\\
\textit{Construction in Theorem 5.1}. The formula $\varPhi_1$ in part 1 is exactly of the form $\varPsi_1 \land \varPsi_2$ defined in \cref{eq:sap}. It can be written in the following way:
\begin{equation}\label{formula_und_2}
	\begin{aligned}
		\varPhi_1 \; = &\; \forallx x_1\forallx x_2\forallx x_3\forallx x_4 \foralla \alpha_1\foralla \alpha_2. \;(x_1 \hookrightarrow x_3 \circ \alpha_1 * x_2 \hookrightarrow x_4 \circ \alpha_2 => x_3 != x_4) \land \forallx x_1\foralla \alpha.\; \lnot (x_1 \hookrightarrow x \circ \alpha).
	\end{aligned}
\end{equation}

The formula $\varPhi_2$ in part 2 can be written in the following way.
\begin{equation}\label{formula_und_3}
	\begin{aligned}
		\varPhi_2 \quad=\quad &\existsx\xz \existsx\xl \existsx\xcf \existsx\xcs \existsx\xf \existsx\xz' \existsx\xl' \existsx\xcf' \existsx\xcs' \existsa\alpha_{c^1}' \existsa\alpha_{c^2}'.\\
		& \mathtt{InitState}(\xz, \xl, \xcf, \xcs, \xf) * \mathtt{FinState}(\xz', \xl', \xcf', \xcs', \alpha_{c^1}', \alpha_{c^2}').
	\end{aligned}
\end{equation}

The predicate $\mathtt{InitState}(\xz, \xl, \xcf, \xcs, \xf)$ encodes the initial state $(I_0, c_0^1, c_0^2) = (1, 0, 0)$ (which corresponds to slave nodes on three master nodes $\xl, \xcf, \xcs$ in Fig. (\ref{fig:reduction}) ), and \\ $\mathtt{FinState}(\xz', \xl', \xcf', \xcs', \alpha_{c^1}', \alpha_{c^2}')$ encodes the final state $(n, c_m^1, c_m^2)$. The state $(n, c_m^1, c_m^2)$ corresponds to slave nodes on three master nodes $\xl', \xcf', \xcs'$ in Fig. (\ref{fig:reduction}). These two predicates are defined as follows:
\begin{equation}\label{eq:init_fin_state}
	\begin{aligned}
		&\mathtt{InitState}(\xz,\xl,\xcf,\xcs, \xf) \\
		\overset{\triangle}{=} \quad& \xz \hookrightarrow \xl \circ \varepsilon * \xl \hookrightarrow \xcf \circ \nil * \xcf \hookrightarrow \xcs \circ \nil * \xcs \hookrightarrow \xf \circ \nil, \\
		& \\
		& \mathtt{FinState}(\xz',\xl',\xcf',\xcs', \alpha_{c^1}', \alpha_{c^2}') \\
		\overset{\triangle}{=} \quad& \xz'\hookrightarrow \xl' \circ \varepsilon * \xl' \hookrightarrow \xcf' \circ \nil^n *  \xcf' \hookrightarrow \xcs' \circ \alpha_{c^1}' \circ \nil * \xcs' \hookrightarrow \varepsilon \circ \alpha_{c^2}' \circ \nil.
	\end{aligned}
\end{equation}
\vspace{0.5em}

Note that $n$ is fixed when the two-counter Minsky machine is given. 

Before constructing the formula $\varPhi_3$, we need to formalize some additional properties shown below. 

The sequence $\alpha$ is of the form $\nil^*$:
\begin{align}\label{eq:ini}
	\mathtt{ini}(\alpha) \overset{\triangle}{=} \nil \circ \alpha == \alpha \circ \nil.
\end{align}

The sequence $\alpha_1$ of the form $\nil^*$ is one element longer than sequence $\alpha_2$ of the same form: 
\begin{align*}
	\alpha_1 == \alpha_2 \circ \nil. 
\end{align*}


With the idea in $\varPsi_4$, we can construct $\varPhi_3$ as follows:
\begin{equation}\label{formula_und_4}
	\begin{aligned}
		\varPhi_3 \quad=\quad& \forallx\xz \forallx\xl \forallx\xcf \forallx\xcs \forallx\xz'.\; \existsx\xl' \existsx\xcf' \existsx\xcs' \exists\xf' \existsa\al \existsa\acf \existsa\acs \existsa\al' \existsa\acf' \existsa\acs'. \\
		& \bigl(\mathtt{CurrState}(\xz,\xl,\xcf,\xcs,\xz', \al,\acf,\acs) \land \lnot \mathtt{FinState}(\xz,\xl,\xcf,\xcs, \alpha_{c^1}, \alpha_{c^2})\bigr) \\
		& => \mathtt{NextState}(\xz,\xl,\xcf,\xcs,\xz',\xl',\xcf',\xcs', \xz'',\al,\acf,\acs,\al',\acf',\acs').
	\end{aligned}
\end{equation}

In $\varPhi_3$, the predicate $\mathtt{CurrState}$ represents the current state beginning from the master node which has no slave nodes, $\mathtt{FinState}$ represents the final state which has been defined in \Cref{eq:init_fin_state}, and $\mathtt{NextState}$ represents the transition from one state to the other. The formula $\varPhi_3$ encodes the following fact: for each current state except for the final state, there exists a next state which can be transitioned from the current state. The predicates $\mathtt{CurrState}$ and $\mathtt{NextState}$ are defined as follows:
\begin{align*}
	& \mathtt{CurrState}(\xz,\xl,\xcf,\xcs,\xz', \al,\acf,\acs)  \\ 
	\overset{\triangle}{=}\quad & \xz \hookrightarrow \xl \circ \varepsilon * \xl \hookrightarrow \xcf \circ \al * \xcf \hookrightarrow \xcs \circ \acf \circ \nil * \xcs \hookrightarrow \xz' \circ \acs \circ \nil, \\
	& \\
	& \mathtt{NextState}(\xz,\xl,\xcf,\xcs,\xz',\xl',\xcf', \xcs',\xz'', \al,\acf,\acs,\al',\acf',\acs') \\
	\overset{\triangle}{=}\quad & \mathtt{Init}(\al,\acf,\acs,\al',\acf',\acs') \land \mathtt{CurrState}(\xz',\xl',\xcf',\xcs',\xz'',\al',\acf',\acs') \land \mathtt{Next}(\al,\acf,\acs,\al',\acf',\acs'), 
\end{align*}
where the predicate $\mathtt{Init}$ is defined as follows. The predicate $\mathtt{ini}$ has been defined in \Cref{eq:ini}.
\begin{align*}
	& \mathtt{Init}(\al,\acf,\acs,\al',\acf',\acs') \\
	\quad\overset{\triangle}{=}\quad &
	\mathtt{ini}(\al) \land \mathtt{ini}(\acf) \land \mathtt{ini}(\acs) \land \mathtt{ini}(\al') \land \mathtt{ini}(\acf') \land \mathtt{ini}(\acs').
\end{align*}

The predicate $\mathtt{Next}$ is defined as follows:
\begin{align*}
	& \mathtt{Next}(\al,\acf,\acs,\al',\acf',\acs') \\
	\overset{\triangle}{=}\quad &
	\bigwedge_{(k_0,c_j,c_j',k) \in \mathcal{I}_1} \Bigl(\al == \nil^{k_0} \land \acj' == \acj \circ \nil \land \acjp' == \acjp \land \al' == \nil^k \Bigr) \\ 
	\land &\; \bigwedge_{(k_0,c_j,c_j',k_1,k_2) \in \mathcal{I}_2} \Bigl(\al == \nil^{k_0} \; \land \bigl(\acj == \nil => (\acf' == \acf \land \acs' == \acs \land \al' == \nil^{k_1})\bigr) \\
	& \hspace{8em} \land \bigl(\lnot (\acj == \nil) => (\acj' \circ \nil == \acj \land \acjp' == \acjp \land \al' == \nil^{k_2})\bigr)\Bigr).
\end{align*}

In the formula $\mathtt{Next}$, the set $\mathcal{I}_1$ and $\mathcal{I}_2$ have been defined in \Cref{def_Minsky}. The reduction can be done following \Cref{formula_und_1,formula_und_2,formula_und_3,formula_und_4}.
$\hfill \square$

\end{document}